\documentclass[preprint2]{aastex}
\usepackage{emulateapj5}
\usepackage{onecolfloat5}
\slugcomment{Submitted to the Astronomical Journal, 6 December 2001;  
        accepted 10 April 2002.}

\def\eg{e.g.}
\def\ie{i.e.}
\def\[{[}
\def\]{]}
\def\<{$<$}
\def\>{$>$}
\def\etal{{\it et~al.\/}\ }
\newcommand{\minusone}{$^{-1}$} 
\newcommand{\LOH}{$L_{OH}$}
\newcommand{\LFIR}{$L_{FIR}$}
\newcommand{\I}{\protect\small I \normalsize $\!\!$}
\newcommand{\II}{\protect\small II \normalsize $\!\!$}
\newcommand{\HI}{\mbox{\rm H\I}\ }

\newcommand{\HII}{\mbox{\rm H\II}\ }

\shorttitle{OH Megamasers III:  The Complete Survey }
\shortauthors{Darling \& Giovanelli}

\begin{document}
\twocolumn[%%% Begin front material
\title{A Search for OH Megamasers at $z > 0.1$. \ \ III.\  The Complete Survey}
\author{Jeremy Darling \& Riccardo Giovanelli}
\affil{Department of Astronomy and National Astronomy and Ionosphere Center, 
	Cornell University, Ithaca,  NY  14853;
        darling@astro.cornell.edu; riccardo@astro.cornell.edu}

\begin{abstract}
We present the final results from the Arecibo Observatory OH megamaser survey.
We discuss in detail the properties of the remaining 18 OH megamasers
detected in the survey, including 3 redetections.
We place upper limits on the OH emission from 85 nondetections and 
examine the properties of 25 ambiguous cases for which the presence or
absence of OH emission could not be determined.  
The complete survey has discovered 50 new OH megamasers (OHMs) in 
(ultra)luminous infrared galaxies (\[U\]LIRGs)
which doubles the sample of known OHMs and increases the 
sample at $z>0.1$ sevenfold.  
The Arecibo OH megamaser survey indicates that the OHM fraction in 
LIRGs is an increasing function of the far-IR luminosity
(\LFIR) and far-IR color, reaching a fraction of roughly one third in the
warmest ULIRGs.  Significant relationships between OHMs and their hosts
are few, primarily due to a mismatch in size scales of measured properties
and an intrinsic scatter in OHM properties roughly equal to the span of the
dataset.  We investigate relationships between OHMs and their hosts 
with a variety of statistical tools including survival analysis, 
partial correlation coefficients, and a principal component analysis.
There is no apparent OH megamaser ``fundamental plane.''  We compile data 
on all previously known OHMs and evaluate the possible mechanisms 
and relationships responsible for OHM production in merging systems.   
The OH-FIR relationship is reexamined using the doubled OHM sample
and found to be significantly flatter than previously thought:
$L_{OH}\propto L_{FIR}^{1.2\pm0.1}$.  This near-linear dependence
suggests a mixture of saturated and unsaturated masers, either within 
individual galaxies or across the sample.  
\end{abstract}
]%%% End front material

\section{Introduction}
We have used the Arecibo telescope\footnote{The Arecibo
Observatory is part of the National Astronomy and Ionosphere Center, which 
is operated by Cornell University under a cooperative agreement with the
National Science Foundation.} to conduct a carefully designed OH 
megamaser (OHM) survey with the primary purpose of building the foundation
required to employ OHMs as tracers of major galaxy mergers, dust-obscured 
starburst nuclei, and the formation of binary supermassive black holes 
spanning the epoch of galaxy evolution.  Results of the survey 
have been presented in installments in Darling \& Giovanelli (2000; hereafter
Paper I) and Darling \& Giovanelli (2001; hereafter Paper II). 
A detailed presentation of the survey selection
criteria is given in Paper I\@.  
Papers I and II presented 35 OHMs, one OH absorber, 
and 160 nondetections, accounting for two-thirds of the survey.  Here
we present the remaining candidates, including
18 OHMs (14 new detections, three redetections, and one new OHM which 
falls outside
of the survey parameter space), one OH absorber, 85 nondetections, and
25 ambiguous cases, and we summarize the results of the entire survey which 
doubles the sample of known OHMs and produces a sevenfold increase in those
at $z>0.1$.  

Darling \& Giovanelli (2002a; hereafter Paper IV) present the derivation of an 
OHM luminosity function and predicts the detectability and areal 
abundance of OHMs as a function of redshift for a number of galaxy 
merger evolution scenarios.  
Darling \& Giovanelli (2002b; hereafter Paper V) present optical
spectroscopy of OHM hosts and a study of OHM environments, lifetimes, and
engines.  

This paper (Paper III) is dedicated to releasing the 
remainder of the new OH megamaser detections in the complete 
Arecibo survey.  We review the candidate selection method in \S 
\ref{sec:selection} and the observation and data reduction methods in 
\S \ref{survey:obs}.  Section \ref{sec:results} presents the results of
the survey including an estimate of the completeness of OHM detections in 
the survey sample.  
We present a compilation of the OHMs known prior to the Arecibo survey, 
investigate the types of merging systems most likely to produce OHMs, and 
analyze the relationships between OHMs and their hosts in \S \ref{sec:hosts}.  

This study parameterizes the Hubble constant as $H_\circ = 75\ h_{75}$
km s\minusone\ Mpc\minusone, assumes $q_\circ = 0$, and 
uses $D_L = (v_{CMB} / H_\circ)(1+0.5\,z_{CMB})$ to 
compute luminosity distances from $v_{CMB}$, the radial velocity of a
radiation source in the cosmic microwave
background (CMB) rest frame.  Line luminosities are always computed
under the assumption of isotropic emission.

\section{Candidate Selection\label{sec:selection}}

For the Arecibo OH megamaser survey, candidates were selected from the 
Point Source Catalog redshift survey (PSCz; Saunders \etal 2000), 
supplemented by the NASA/IPAC Extragalactic 
Database\footnote{The NASA/IPAC Extragalactic Database (NED) is operated 
by the Jet Propulsion 
Laboratory, California Institute of Technology, under contract with the 
National Aeronautics and Space Administration.}.  
The PSCz catalog is a flux-limited ({\it IRAS} $f_{60\mu m} > 0.6$ Jy)
redshift survey of 15,000 {\it IRAS} galaxies over $84\%$ of the sky
(see Saunders et al. 2000).
We select {\it IRAS} sources which are in the Arecibo sky ($0^\circ <
\delta <
37^\circ$), were detected at 60 $\mu$m, and have $0.1\leq  z \leq 0.45$.  
The lower redshift bound is set to avoid local radio frequency interference
(RFI), while the upper bound is set by the bandpass of the wide L-band receiver
at Arecibo, although an effective upper bound is imposed around $z=0.23$
by the RFI environment, as discussed in 
\S \ref{subsec:ambig}.  
No constraints are placed on far-infrared (FIR) colors or luminosity.  The 
redshift requirement limits the number of candidates in the Arecibo sky 
to 311.  The condition that
candidates have $z>0.1$ automatically selects (ultra)luminous infrared 
galaxies ([U]LIRGs) if they are included
in the PSCz.  The strong influence of \LFIR\ and FIR color on OHM fraction 
in LIRGs is the primary reason for our high detection
rate compared to previous surveys (\eg\ Staveley-Smith \etal 1992; 
Baan, Haschick, \& Henkel 1992).  Baan (1991), Staveley-Smith \etal (1992), 
Baan, Haschick, \& Henkel (1992), and others have noted the dependence of OHM
fraction on the FIR luminosity and color, and these relationships are 
reexamined in \S \ref{sec:ohfrac}.

\section{Observations and Data Reduction\label{survey:obs}}

The upgraded Arecibo radio telescope offers new opportunities for the
detection of OHMs, due to its improved sensitivity, frequency agility, and
instantaneous spectral coverage.  Its large collecting area 
makes it ideal for a survey of spectral lines at the upper end of the 
redshift range of the known OHM sample ($0.1 \leq z \leq 0.3$).  
Detection of OH emission lines
is generally possible in a 4-minute integration, even at $z \simeq 0.2$.  
In roughly 200 hours of telescope time, we were able to observe about 
300 OHM candidates and double the sample of known OHMs.

As described in Paper II, 
observations at Arecibo with the L-band receiver were performed 
by nodding on- and off-source, each for a 4 minute total integration, followed
by firing a noise diode.  Spectra were recorded in 1 s intervals 
to facilitate radio frequency interference 
flagging and excision in the time-frequency domain.  Note that 
early observations reported in Paper I were 
sampled only every 6 s.
Data were recorded with 9-level sampling in 2 polarizations
of 1024 channels each, spanning a 25 MHz bandpass centered on redshifted
1666.3804 MHz (the mean of the 1667.359 and 1665.4018 MHz OH lines).  

All reduction and analysis of these fast-sampled data was performed with the 
AIPS++ software package\footnote{The AIPS++ (Astronomical Information 
Processing System) is a 
product of the AIPS++ Consortium. AIPS++ is freely available for use under 
the Gnu Public License. Further information may be obtained from 
http://aips2.nrao.edu.} 
using home-grown routines for single dish reduction which are described in
detail in Paper II.  We included 
in the reduction pipeline RFI flagging routines designed to identify two
types of RFI observed at Arecibo:  strong features more
than $3\sigma$ above the time-domain noise and spectrally broad, 
low-level RFI which is time-variable.  
A reliable estimate of
the weight, or effective integration time, on a single channel is the 
total number of records used in forming the time average, excluding flagged 
records.  Each spectral channel in a 
calibrated time-averaged spectrum may have a different effective integration
time and hence a different effective intrinsic radiometer noise level.
We include a normalized weights spectrum with each OHM spectrum  
in Figure \ref{spectra}.  Depressions in the weights spectra 
indicate the presence of RFI, which may or
may not have been completely flagged and removed from the final spectra.  
Spectral channels with low weights should thus be treated with skepticism.
The frequency resolution after hanning smoothing is 49 kHz (10 km s\minusone\
at $z=0.1$), and the uncertainty in the absolute flux scale is $8\%$.

\section{Survey Results\label{sec:results}}

The survey selection criteria identify 312 candidate OHMs in the
PSCz.  One of these, {\it IRAS} 10232+1258, has an incorrect redshift in the
catalog, and the correct redshift places it below the $z>0.1$ selection 
threshold ($z=0.0325$ ; Haynes \etal 1997).  Hence, of the 311 remaining
candidates, 
52 are OHMs (3 were previously identified; 15 are new detections reported
in this paper), 1 is an OH absorber (reported here),
233 do not show OH emission lines down to the detection threshold (85 of 
these are reported in this paper), and 
25 remain ambiguous (reported here) mostly due to anthropogenic RFI and 
Galactic \HI ``RFI\@.'' The latter appears at $z=0.174$ (1420 MHz), 
where strong Galactic 
\HI emission interferes with the detection of comparatively weak OH lines.  
Overall, we find one OH megamaser in every 5.5 candidates.  Further 
discussion of the OHM fraction in mergers is presented in \S \ref{sec:hosts}.
Three objects deserve special mention:  {\it IRAS} F13451+1232 has a tentative
OH detection (Dickey \etal 1990), but could not be confirmed in this survey
due to its strong radio continuum (see \S \ref{subsec:ambig}) and is listed 
with the ambiguous candidates (it is not included in detection statistics and
is not counted as an OHM in the survey); 
{\it IRAS} F11180+1623 is not included in the 
PSCz sample, but is a new OHM detection and is listed in the OH 
detection tables and spectra but is not included in the survey statistics 
and analysis; and  {\it IRAS} F19154+2704 is not included in the 
PSCz sample, but is a new OH absorber listed in Paper I
which is not included in the survey statistics and analysis.

\subsection{Nondetections\label{survey:nond}}

Tables \ref{nondetectFIR} and \ref{nondetectOH} list respectively 
the optical/FIR and radio properties of the 85 OH non-detections in the 
last third of the survey.  Table \ref{nondetectFIR} lists the optical redshifts
and FIR properties of the non-detections in the following format:
Column (1) lists the  {\it IRAS} Faint Source Catalog (FSC) name.  
Columns (2) and (3) list the  source coordinates (epoch B1950.0) 
from the FSC, or the Point Source Catalog (PSC) if unavailable in the FSC.  
Columns (4), (5) and (6) list the heliocentric optical redshift, 
reference, and corresponding velocity.  Uncertainties in velocities are 
listed whenever they are available.  
Column (7) lists the cosmic microwave background rest-frame velocity.  This is
computed from the heliocentric velocity using the solar motion with respect
to the CMB measured by Lineweaver \etal (1996):  
$v_\odot = 368.7 \pm 2.5$ km s\minusone\
towards $(l,b) = (264\fdg31 \pm 0\fdg16 , 48\fdg05 \pm 0\fdg09)$.
Column (8) lists the luminosity distance computed from $v_{CMB}$ via 
$D_L = (v_{CMB} / H_\circ)(1+0.5z_{CMB})$, assuming $q_\circ = 0$.  
Columns  (9) and (10) list the {\it IRAS} 60 and 100 $\mu$m flux 
densities in Jy.  FSC flux densities are listed whenever they are 
available.  Otherwise, PSC flux densities
are used.  Uncertainties refer to the last digits of each measure, and upper 
limits on 100 $\mu$m flux densities are indicated by a ``less-than'' symbol. 
Column (11) lists the logarithm of the far-infrared luminosity in units
of $h_{75}^{-2} L_\odot$.  
\LFIR\ is computed following the prescription of Fullmer \& Lonsdale (1989):  
\LFIR$ = 3.96\times 10^5 D_L^2 (2.58 f_{60} + f_{100})$, 
where $f_{60}$ and $f_{100}$ are the 60 and 100 
$\mu$m flux densities expressed in Jy, $D_L$ is in $h_{75}^{-1}$ Mpc, 
and \LFIR\ is in units of $h_{75}^{-2} L_\odot$.  
If $f_{100}$ is only available as an upper limit, the permitted range
of \LFIR\ is listed.  The lower bound on \LFIR\ is computed for $f_{100}=0$ Jy,
and the upper bound is computed with $f_{100}$ set equal to its upper limit.
The uncertainties in $D_L$ and in the {\it IRAS} flux densities 
typically produce an uncertainty in $\log L_{FIR}$ of $0.03$.  

Table \ref{nondetectOH} lists the 1.4 GHz 
flux density and the limits on OH emission of the non-detections in 
the following format:
Column (1) lists the  {\it IRAS} FSC name, as in Table \ref{nondetectFIR}.
Column (2) lists the heliocentric optical redshift, as in 
Table \ref{nondetectFIR}.
Column (3) lists $\log$ \LFIR, as in Table \ref{nondetectFIR}.
Column (4) lists the predicted isotropic OH line luminosity, 
$\log L_{OH}^{pred}$,
based on the Malmquist bias-corrected $L_{OH}$-\LFIR\ relation 
determined by Kandalian (1996) from 49 OHMs:  
$\log L_{OH} = (1.38\pm0.14) \log L_{FIR} - (14.02\pm 1.66)$ 
(see \S \ref{sec:relation}).
Column (5) lists the upper limit on the isotropic OH line luminosity, 
$\log L_{OH}^{max}$.  
The upper limits on \LOH\ are computed from the RMS noise of the non-detection
spectrum assuming a ``boxcar'' line profile of rest frame width 
$\Delta v = 150$ km s\minusone\ and height 1.5$\sigma$: 
\begin{equation}
L_{OH}^{max} = 4 \pi D_L^2\ 1.5 \sigma \left({\Delta v\over c}\right) 
\left({\nu_\circ \over 1+z}\right).   
\end{equation}
The assumed rest frame width $\Delta v = 150$ km
s\minusone\ is the average FWHM of the 1667 MHz line of the known OHM sample.
Column (6) lists the on-source integration time, in minutes.  
Column (7) lists the RMS noise values in flat regions of the 
non-detection baselines, in mJy, after spectra were hanning smoothed 
to a spectral resolution of 49 kHz.
Column (8) lists 1.4 GHz continuum fluxes, from the NRAO 
VLA Sky Survey (NVSS; Condon et al. 1998).  If no continuum source lies within 
30\arcsec\ of the {\it IRAS}
coordinates, an upper limit of 5.0 mJy is listed.  
Column (9) lists the optical spectroscopic classification, if available.  
Codes used are:  ``S2'' = Seyfert type 2;
``S1'' = Seyfert type 1;  ``H'' = \HII region (starburst);  
and ``L'' = low-ionization emission region (LINER).  References
for the classifications are listed in parentheses and included at the 
bottom of the Table.
Column (10) lists source notes, which are given at the foot of the table.

We can predict the expected isotropic OH line luminosity, $L_{OH}^{pred}$, 
for the OHM candidates based on the 
$L_{OH}$-\LFIR\ relation determined by Kandalian (1996; see \S 
\ref{sec:relation}) 
and compare this figure to upper limits on the OH emission derived
from observations, $L_{OH}^{max}$, for a rough measure of the 
confidence of the non-detections.  Note, however, that 
the scatter in the $L_{OH}$--$L_{FIR}$ relation is quite large:  roughly 
half an order of magnitude in \LFIR\ and one order of magnitude in \LOH\ (see
Kandalian 1996).  Among the non-detections, 34 out of 233 galaxies have 
$L_{OH}^{pred} < L_{OH}^{max}$, indicating that longer integration times 
are needed to unambiguously confirm these non-detections, and
33 out of 233 candidates have $L_{OH}^{max}$ within the range of 
$L_{OH}^{pred}$ set by an upper limit on $f_{100}$.  
Integration times were a compromise between efficient use of telescope
time and the requirement for a meaningful upper limit on \LOH\ for
non-detections.

\subsubsection{Notes on Nondetections}

\noindent{\bf 12514+1027: }
There are two redshifts for this object in the literature:  
0.30 (Wilman \etal 1998) and 0.3189 (PSCz).  We have 
coverage of both redshifts and find no clear OH emission, 
but there are regions in the bandpass between the two frequencies  
which are unobservable due to RFI\@.  We use the PSCz redshift as the
fiducial for computing luminosities and limits on the OH emission of
this object.

\subsection{Ambiguous Candidates\label{subsec:ambig}}
Despite careful observations and RFI mitigation measures, there remain 
25 OHM candidates in the survey for which no OH measurement could be
made.  One of these, {\it IRAS} 13451+1232, is likely to be an OHM
(Dickey \etal 1990).  There are two factors which frustrate weak spectral
line observations:  RFI and strong radio continuum in the beam.  

RFI can be
anthropogenic or Galactic \HI which makes OH undetectable at 
velocities around 52000 km s\minusone.  The effect of RFI is to constrain
the survey sample in redshift.  A group of candidates near $z=0.174$ is
excluded from the sample by Galactic \HI, and most candidates above 
$z=0.23$ are not observable except in small RFI-free windows.  The 
exclusion of candidates by the Galactic \HI will not significantly bias
the survey, but the irregular redshift coverage above 
$z=0.23$ will 
produce a bias.  The best solution to this bias would be to impose
an upper cutoff in redshift for the survey.

Strong radio continuum sources 
(including the Sun) produce standing waves between the instrument platform
and the primary reflector at Arecibo, making complicated baseline structures
which obliterate the signatures of weak spectral lines.  This may be a 
significant
source of bias for the survey, because these IR quasars represent a special
population which may become increasingly important at higher redshifts.  
However, there are only 4 strong continuum sources in the survey which are RFI-free, 
representing a small potential contribution to the overall survey statistics.  

Tables \ref{ambig:opt} and \ref{ambig:radio} list, respectively,
the optical/FIR and radio properties of the 25 ambiguous candidates.
The column headings of Table \ref{ambig:opt} are identical to those of Table
\ref{nondetectFIR}, except for the final column which lists notes 
describing the source of ambiguity for each object.  
Table \ref{ambig:radio} lists the 1.4 GHz flux 
density, the optical classification, and notes on each candidate
in the following format:
Column (1) lists the {\it IRAS} FSC name, as in Table \ref{nondetectFIR}.
Column (2) lists the heliocentric optical redshift, as in Table \ref{nondetectFIR}.
Column (3) lists $\log$ \LFIR, as in Table \ref{nondetectFIR}.
Column (4) lists the predicted isotropic OH line luminosity, 
$\log L_{OH}^{pred}$,
based on the Malmquist bias-corrected $L_{OH}$-\LFIR\ relation 
determined by Kandalian (1996; see \S \ref{sec:relation}).
Column (5) lists 1.4 GHz continuum fluxes from the NVSS.  
If no continuum source lies within 
30\arcsec of the {\it IRAS} coordinates, an upper limit of 5.0 mJy 
is listed.  
Column (6) lists the optical spectroscopic classification, if available:
(S1) Seyfert type 1,  
(S2) Seyfert type 2,
(Q) Quasar,
and (L) low-ionization emission region (LINER).  References
for the classifications are shown in parentheses and included at the 
foot of the table.
Column (7) lists source notes, which are given at the foot of the table.

\subsubsection{Notes on Ambiguous Detections}

\noindent{\bf 13451+1232: }  
We attempted to make a re-observation of this tentative OH detection by 
Dickey \etal (1990), but 
the strong radio continuum (5.4 Jy) produced standing waves
in the spectrum and frustrated detection of any OH lines.  
The host of this probable OHM is classified a Seyfert 2 by Kim, 
Veilleux, \& Sanders (1998) and a Seyfert 1.5 by Baan, Salzer, 
\& LeWinter (1998) (the latter indicate that two nuclei were observed, but
only one classification is listed for this object).
Veilleux, Sanders, \& Kim (1997) have detected broad Pa$\alpha$ 
emission from this merger, indicating that it contains a buried quasar.  
In fact, many groups refer to this object as an ``IR quasar''.
This is a double nucleus system, with nuclear separation of 
3\arcsec\ (7.1 kpc) and ample molecular gas:  $\log M(H_2) = 10.78$ 
(Scoville \etal 2000). 
%DSS image shows a large symmetrical galaxy with a hint of extension 
%in the NW.  Abundant HST data available.  Clearly a double nucleus in an 
%extended envelope.  Cluster???

\begin{figure*}[!ht]
%\plotfiddle{Darling.fig1a.epsi}{7in}{0}{100}{100}{-310}{-155}
%\centerline{\psfig{figure=figures/allspec1.epsi,clip=t,width=6in}}
\epsscale{1.5}
\plotone{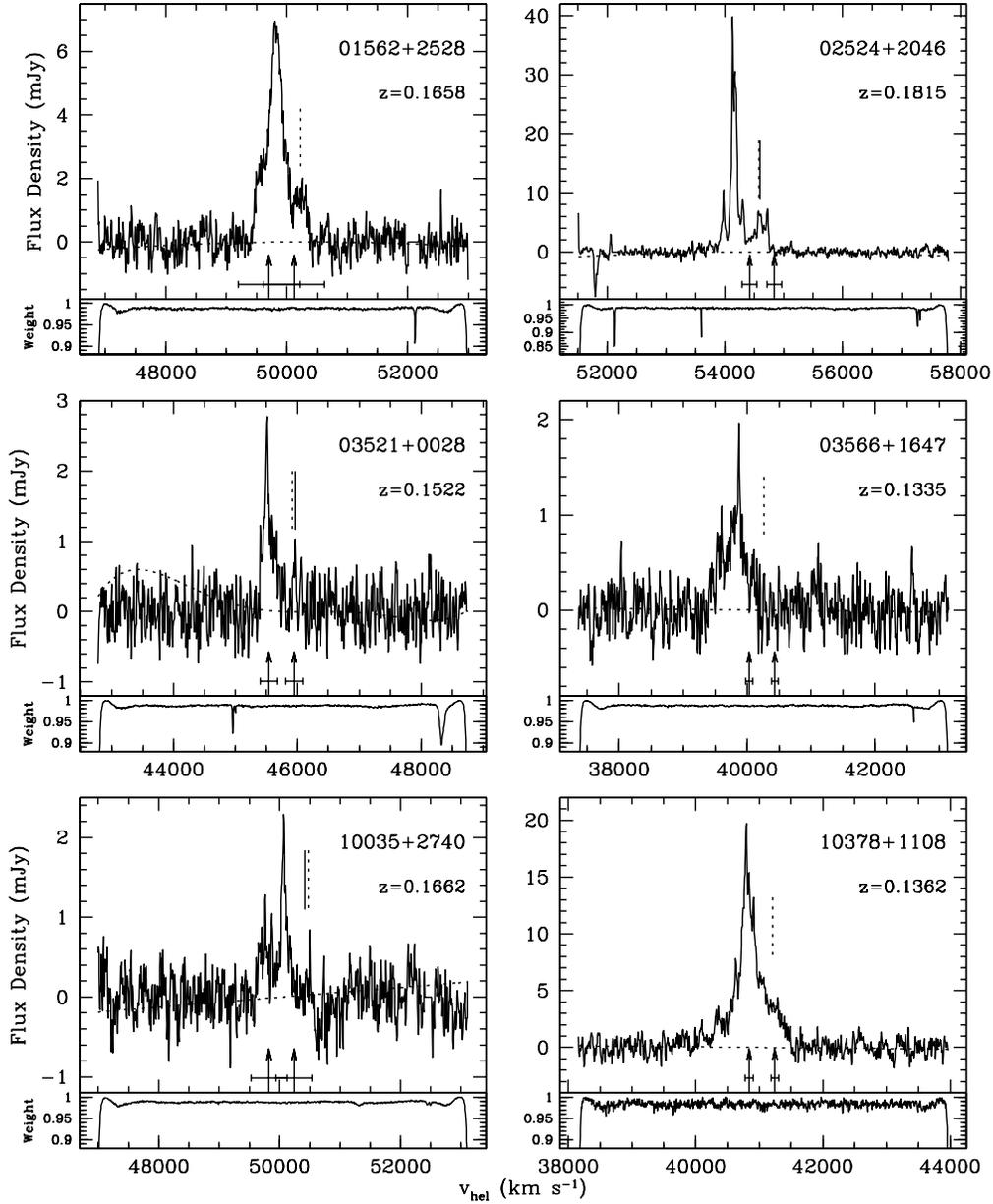}
\caption[OH Megamaser Spectra]{New OH megamasers discovered in (U)LIRGs.  
Abscissae and inset redshifts refer to the optical heliocentric velocity.
Spectra use the 1667.359 MHz line as the rest frequency for the 
velocity scale.  
Arrows indicate the expected velocity of the 1667.359 ({\it left})
and 1665.4018 ({\it right}) MHz lines based on the optical redshift, with 
error bars indicating the uncertainty in the redshift.  Solid vertical
lines indicate the location of the secondary maximum in the autocorrelation
function, and dashed vertical lines indicate the expected position of the 
1665 MHz line, based on the centroid of the 1667 MHz line; a match between
the two indicates a possible detection of the 1665 MHz line.  The 
dotted baselines indicate the shape (but not the absolute magnitude) 
of the baselines subtracted from the calibrated spectra.
The small frame below each spectrum shows the ``weights'' spectrum, indicating
the fractional number of RFI-free records averaged in each channel.
The properties of these megamasers are listed in Tables \ref{detectFIR}
and \ref{detectOH}.\label{spectra}}
\end{figure*} 

\begin{figure*}[!ht]
\figurenum{1}
%\centerline{\psfig{figure=Darling.fig1b.epsi}}
\epsscale{1.5}
\plotone{Darling.fig1b.epsi}
\caption{{\it continued.}\label{spectrab}}
\end{figure*}

\begin{figure*}[!ht]
\figurenum{1}
%\centerline{\psfig{figure=Darling.fig1c.epsi}}
\epsscale{1.5}
\plotone{Darling.fig1c.epsi}
\caption{{\it continued.}\label{spectrac}}
\end{figure*}

\subsection{OH Megamaser Detections}

\begin{figure}[t!]
\epsscale{1}
\plotone{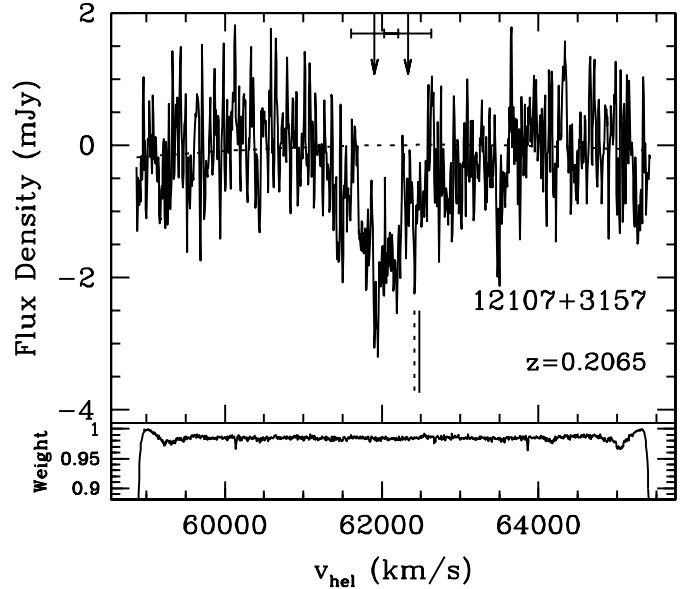}
%\centerline{\psfig{figure=Darling.fig2.epsi}}
\caption[An OH Absorber Discovered in a ULIRG]
{An OH Absorber Discovered in a ULIRG.  This spectrum is presented
in exactly the same manner as the OH megamaser detections (see Figure 
\ref{spectra}).  The properties of this OH absorber is listed
at the end of Tables \ref{detectFIR} and \ref{detectOH}.
\label{absspectrum}}
\end{figure}

Tables \ref{detectFIR} and \ref{detectOH} list respectively the 
optical/FIR and radio properties of the 15 new OHM detections, the 
three redetections, and one OH absorber.  Note that OHM {\it IRAS} F11180+1623
is {\it not} in the PSCz sample, but was observed
along with other OHM candidates not found in the PSCz sample to fill in 
telescope time when local sidereal time coverage of the official sample was 
sparse.  This
detection is not included in any survey statistics or interpretation.
Spectra of the 18 OHMs appear in Figure \ref{spectra}, and spectra of the
OH absorber appears in Figure \ref{absspectrum}.  
The column headings of Table \ref{detectFIR} are identical to those of Table
\ref{nondetectFIR}.  Table \ref{detectOH} lists the OH emission
properties and 1.4 GHz flux density of the OH detections in the following
format:
Column (1) lists the {\it IRAS} FSC name.
Column (2) lists the measured heliocentric velocity of the 1667.359 MHz 
line, defined by the center of the FWHM of the line.  The uncertainty in the
velocity of the line center is estimated assuming an uncertainty of $\pm 1$
channel ($\pm 49$ kHz) on each side of the line.  Although one can generally
determine emission line centers with much higher precision when the shapes
of lines are known, the OHM line profiles shown in Figure \ref{spectra} 
are asymmetric, multi-component, and non-gaussian.  They defy simple 
shape descriptions, so we use this conservative and basic prescription 
to quantify the uncertainty in the line centers.
Column (3) lists the on-source integration time in minutes.  
Column (4) lists the peak flux density of the 1667 MHz OH line in mJy.
Column (5) lists the equivalent width-like measure in MHz.  
$W_{1667}$ is the ratio of the integrated 1667 MHz line flux to its 
peak flux.  Ranges are listed for $W_{1667}$ in cases where the identification
of the 1665 MHz line is unclear, but in many cases the entire emission 
structure is included in $W_{1677}$ as indicated in the discussion of 
each source below.
Column (6) lists the observed FWHM of the 1667 MHz OH line in MHz.
Column (7) lists the rest frame FWHM of the 1667 MHz OH line in km 
s\minusone.   The rest frame width was calculated from the observed width as
$\Delta v_{rest} = c (1+z) (\Delta \nu_{obs} / \nu_\circ)$.
Column (8) lists the hyperfine ratio, defined by $R_H = F_{1667}/F_{1665}$, 
where 
$F_\nu$ is the integrated flux density across the emission line centered on 
$\nu$.  $R_H = 1.8$ in thermodynamic equilibrium.  In many cases, the 1665 MHz
OH line is not apparent, or is blended into the 1667 MHz OH line, and a good
measure of $R_H$ becomes difficult without a model for the line profile.  It is
also not clear that the two lines should have similar profiles, particularly 
if the
lines are aggregates of many emission regions in different saturation states. 
Some spectra allow a lower limit to be placed on $R_H$, indicated by a 
``greater than'' symbol.  
Blended or noisy lines have uncertain values of $R_H$, and are indicated 
by a tilde, but in some cases, separation of the two OH lines is impossible
and no value is listed for $R_H$.  
Column (9) lists the logarithm of the FIR luminosity, as in Table 
\ref{detectFIR}.
Column (10) lists the predicted OH luminosity, $\log L_{OH}^{pred}$, as in 
Table \ref{nondetectOH}.
Column (11) lists the logarithm of the measured isotropic OH line luminosity, 
which includes the 
integrated flux density of both the 1667.359 and the 1665.4018 MHz 
lines.
Note that $L_{OH}^{pred}$ is generally 
less than the actual \LOH\ detected (46 out of 53 detections).
Column (12) lists the 1.4 GHz continuum fluxes from the NVSS.
If no continuum source lies within 30\arcsec\ of the {\it IRAS} 
coordinates, an upper limit of 5.0 mJy is listed.  

The spectra of the OH detections are presented in Figures \ref{spectra} 
and \ref{absspectrum}.  
The abscissae and inset redshifts refer to the optical heliocentric 
velocity, and the arrows indicate the expected velocity of the 1667.359 
({\it left}) and 1665.4018 ({\it right}) MHz lines based on the optical 
redshift, with error bars indicating
the uncertainty in the redshift.  The spectra refer to  1667.359 MHz as 
the rest
frequency for the velocity scale.  Spectra have had the dotted baselines
subtracted, and the baselines have been shifted in 
absolute flux density such that the central channel has value zero.  
The small frame below each spectrum shows a weights spectrum, indicating
the fractional number of records used to form the final spectrum after
the RFI rejection procedure (see \S \ref{survey:obs} and Paper II).  
Channels with weights close to unity are ``good''
channels, whereas channels with lower than average weight are influenced by
time-variable RFI and are thus suspect.  The weights spectra are presented
to indicate confidence in various spectral features, but note that often
the RFI rejection procedure does a good job of cleaning channels and that
channels with $\sim10\%$ rejected records may be completely reliable (this
is, after all, the point of the RFI cleaning procedure).

In order to quantitatively identify dubious 1665 MHz OH line
detections, we compute the autocorrelation function (ACF) of each spectrum and 
locate the secondary peak (the primary peak corresponds to zero offset, or
perfect correlation).   Any correspondence of features between the two main 
OH lines will enhance the second autocorrelation peak and allow us to 
unambiguously identify 1665 MHz lines based not strictly on spectral 
location and peak flux, but on line shape as well.
The secondary peak in the ACF of each spectrum, when present, 
is indicated by a small solid line over the spectra in Figure \ref{spectra}.  
We expect 
the offset of the secondary peak to be equal to the separation
of the two main OH lines, properly redshifted:  (1.9572 MHz)$/(1+z)$.  The 
{\it expected} location of the secondary ACF peak is indicated
in Figure \ref{spectra} by a small dashed line over each spectrum.  
Both the expected and actual secondary peak positions are plotted 
offset with respect to the center of the 1667 MHz line, as defined by the 
center of the FWHM, rather than the peak flux.  

We examined the Digitized Sky Survey\footnote{Based on 
photographic data obtained using Oschin Schmidt Telescope on Palomar 
Mountain. The Palomar Observatory Sky Survey was funded by the National 
Geographic Society. The Oschin Schmidt Telescope is operated by the 
California Institute of Technology and Palomar Observatory. The plates 
were processed into the present compressed digital format with their
permission. The Digitized Sky Survey was produced at the Space Telescope 
Science Institute under U.S. Government grant NAG W-2166.}
(DSS) images of each new OH detection.   The OHM hosts 
are generally faint, unresolved, and unremarkable in the DSS unless otherwise
noted in the discussion of individual sources below.  We also performed
an exhaustive literature search for each new OHM, and searched the Hubble
Space Telescope (HST) archives for fields containing OHM hosts.  All 
relevant data are included in the discussions below.  The weights spectra
are generally clean across the OH line profiles, unless specifically noted.
Note that the sampling rate for the first OHM detections in the survey 
(Paper I) was
too slow to perform the RFI cleaning procedure discussed in Paper II,
so these OHMs do not have weights spectra available. 
Included in the discussion below are the optical spectral types of OHM hosts 
from the literature and from observations presented in Paper V\@.
We make some observations and measurements specific to individual OH detections
as follows.

\subsubsection{Notes on OH Megamasers}

\noindent{\bf 01562+2528: }
This OHM shows broad, blended lines, which are easily identified, despite the
highly uncertain optical redshift.  Although the ACF shows no second peak,
the 1665 MHz line is quite prominent.  Galactic \HI has been masked at
52200 km s\minusone.  
The DSS image of the host of this OHM shows a 
multi-component object with possible extended emission.  
Observations at the Palomar 5 m telescope confirm this with the detection
of two nuclei (Paper V).  We classify the northeast 
nucleus of this OHM host a starburst.  The southwest nucleus lacked sufficient
signal for optical classification, but a redshift was measured (see 
Paper V).  The velocity of the NE nucleus is 
consistent with the OH redshift but the SW nucleus is not.
%F01562 has MWHI masked out (shows in weights).  ACF is very smooth (no 2nd pk).  
%Good match to (uncertain) optical redshift.  1667 ID is easy although ACF method
%fails (linewidth).  Clean weights.  DSS shows multi-component object
%w/ possible extended emission.  
%No HST

\noindent{\bf 02524+2046: }
This OHM has the most unusual spectrum of the survey sample.  The peak flux
density is 40 mJy; RFI and receiver considerations aside, such an OHM would be 
detectable beyond $z=1$ at Arecibo in a short integration time.
The emission lines are 
strong, narrow, and show extremely good correspondence between 1667 and 1665 lines, 
including a blue tail on each complex.  The correspondence is so good that we
can measure the hyperfine ratio for each component of the line profile.
From high velocity to low, we find $R_H$ = 1.40, 5.63, 1.88 (Table \ref{detectOH}
lists the hyperfine ratio of the combined emission: $R_H=3.2$).  Galactic \HI 
has been masked in the spectrum, and is accompanied by a dip in the weights spectrum.
There is also an interesting feature at 51800 km s\minusone\ which is likely to 
be \HI in the off-source position (5$^m$ later than the on-source position) 
with a heliocentric velocity of about $-300$ km s\minusone.  This is probably
associated with the Magellanic stream.  
We classify the host of this OHM a starburst (Paper V).
%flux(1665,1667) = (1.46,2.04),(2.35,13.24),(1.34,2.52) and 
%We measure the hyperfine ratio 
%of the three obvious emission lines.  
%Lines are well separated, and quite strong.
%Looks like there are multiple well-defined masing regions (separated by velocity).
%MW HI was masked out, but is still indicated by a dip in the weights spectrum.  This
%is an extremely luminous OHM, weighing in at $\sim3.71$.  Note about abs feature...
%Imperfect match between optical and OH velocities.  ACF and centroid predctions
%for 1665 line are bang-on, and agree with each other.  Uncanny correspondence of
%features.  Lines are narrow, and hyperfine ratio is lowish (compare to overall 
%sample).  
%No obvious optical counterpart.
%  No HST.

\noindent{\bf 03521+0028: }
The optical redshift perfectly matches the 1667 MHz line as well as the marginally
significant 1665 MHz line.  The feature identified as the 1665 MHz line by the
ACF and the position of the 1667 MHz line is a 3.4$\sigma$ detection and is quite
narrow.  The host of this OHM is classified 
in the mid-IR as starburst by Lutz, Veilleux, \& Genzel (1999) and in the optical 
as a LINER by Veilleux, Kim, \& Sanders (1999) and us (Paper V).
K-band imaging by Murphy \etal (1996) gives a nuclear separation of 1.6\arcsec\ or
3.6 kpc.  Solomon \etal (1997) measure a CO(1--0) line width of 150 km 
s\minusone, estimate a H$_2$ mass of 4.3$\times 10^{10} M_\odot$, 
and derive a blackbody radius of 319 pc.  
%F03521 has a very tenuous (non-significant) 1665 line (peak is at 3.4$\sigma$).  
%2nd peak in ACF is very weak and broad (double, in fact).  Clean weights spec.
%Optical redshifts are bang-on!  ACF and centroid agree roughly on the marginal
%feature.  RFI missed the spectral lines.  Was observed at vhel=45541 km/s (change
%and check v,z,L,D in tables) which is different from the PSCz velocity.  
%Currently:  v1667ctr is correct, according to observed frequency.  
%Possible HST images do not include this object.  
%Unresolved DSS image.  Clear optical counterpart.

\noindent{\bf 03566+1647: }
There is no obvious 1665 MHz line in this OHM spectrum, and the ACF shows no 
secondary peak.  We compute an upper limit on the 1665 MHz line flux assuming
a 1$\sigma$ line of width equal to the 1667 MHz line to obtain a lower limit
on the hyperfine ratio:  $R_H \gtrsim 9.6$.  The optical redshift does not
correspond to the peak OH emission.  
We classify the host of this OHM a Seyfert 2 (Paper V).
%F03566 has no 2nd pk in ACF, and no obvious 1665 line.  clean weights spec.  Used
%whole line as 1667 MHz for the equiv width measure.  Optical z is off the peak
%of OH emission.  No ACF.  No feature at centroid prediction for 1665.  Will need
%method for $R_H$ and range for $W_{1667}$. RMS = 0.203 mJy across expected 1665 line.
%No significant area.  Rough $R_H \gtrsim 9.6$.
%No HST.  DSS img looks extended.  

\noindent{\bf 10035+2740: } 
This OHM has a number of sharp lines, and it is unclear if the 1665 MHz 
line is present (there is a narrow 4$\sigma$ line somewhat above the expected 
position which is included in the estimate of $R_H$).  
Galactic \HI has been masked in the final spectrum.  
The spectrum is somewhat noisy, and it is unclear if the multiple peaks are 
OH or noise; there are two broad lines and smaller spikes on the less 
prominent line.  The weights spectrum is very clean.  
%Unresolved in DSS.  No HST.

\noindent{\bf 10378+1108: } 
% Oct99 observations
This OHM is a re-detection, and was discovered by Kazes \& Baan (1991).
The measured properties listed in Table \ref{detectOH} are consistent 
with those measured by Kazes \& Baan (1991), and the spectra are similar,
but comparison is difficult due to very different signal to noise 
observations.  The spectrum in Figure \ref{spectra} shows a strong
line with broad red and blue wings with no clear 1665 MHz line.
The ACF of this spectrum shows no second peak.  K-band imaging by 
Murphy \etal (1996) obtain an upper limit on the nuclear separation (if any)
of this OHM host:  $< 0.5$\arcsec\ or $<1.1$ kpc.  
The host of this OHM is classified a 
LINER by Veilleux, Kim, \& 
Sanders (1999) and by us (Paper V).
%DSS image shows a slightly elongated but otherwise unresolved
%object.  No HST.

\noindent{\bf 11180+1623: } 
{\it Note that this object is NOT in the PSCz sample.}
The emission profile of this OHM has a sharp red falloff and a blue tail.  
The 1665 MHz line identified by the ACF and the 1667 MHz line prediction
has 2.8$\sigma$ significance, but resembles other baseline features.  The
hyperfine ratio measured from this line is thus treated as a lower limit.
Galactic \HI has been masked in the spectrum presented in Figure 
\ref{spectra}.  
%Good baseline, masked MW HI, clean weights.  No HST.
%Unresolved in DSS - may be a faint companion?  The ACF and 1667 predictions
%for 1665 line are coincidental (on top of each other).

\noindent{\bf 12005+0009: } 
This OHM has at least three distinct emission components and a very broad
total spectrum (1081 km s\minusone\ in the rest frame at 10$\%$ of the 
peak flux density).  
The ACF is broad and multiply-peaked, but seems to indicate the secondary
component as the 1665 MHz line, in agreement with the prediction from the
center of the main 1667 MHz line.  
All components are included in the total OH measure, but we tentatively 
identify the secondary component as the 1665 MHz line for a rough 
measure of the hyperfine ratio which excludes the highest velocity 
tertiary component:  $R_H\sim2.0$.  The DSS image of the OHM host is 
extended and irregular.
%F12005 has smooth weights across OH spectrum (not clean everywhere). It has a 
%very 
%broad spectrum (1.487215-1.48186 = 5.355 MHz = 963 km/s at 10$\%$ peak flux), 
%and the 1665 component is anyone's guess.  The 1667 measurement above is also 
%a guess.
%ACF is broad and multiply-peaked.  The 1665 measurement included the secondary
%structure, but not the tertiary one (which IS in the total OH measure).
%Extended, irregular DSS image (looks like a merger).  No HST.

\noindent{\bf 12018+1941: }  
We redetect an OHM first observed by Martin \etal (1988).  
The spectra look similar and all of the measurements presented in Table
\ref{detectOH} are consistent with those of Martin \etal (1988), but the 
uncertainties are high in both detections due to low signal to noise.  
Galactic \HI has been masked in the spectrum presented in Figure 
\ref{spectra}.  The ACF shows a minor second peak, although there is no
significant spectral feature where we expect a 1665 MHz line.  
We compute a lower bound on $R_H$, assuming
a 1$\sigma$ 1665 MHz profile with width equal to the 1667 MHz line:  
$R_H\geq5.6$.
The host of this OHM is classified a LINER by Baan, Salzer, \& LeWinter (1998)
and by Veilleux, Kim, \& Sanders (1999), although the latter group notes
that it could be a Seyfert 2.  
%Edge-like baseline - low below MW HI, high above the OH.  1665 is not obvious.
%ACF is swamped by MW HI.  With MW HI masked, there is a minor second peak.
%RMS in 1665 region is 0.44 mJy.  
%Clean weights.  MW HI in bandpass (no problem).
%DSS image shows an extended but otherwise unresolved object.  No HST.

\noindent{\bf 12162+1047: } 
This OHM has no obvious 1665 MHz line, and mild standing waves in the 
baseline.  We
measure only the main line, since the smaller line on the blue side is similar
in size and shape to noise features in the bandpass.  The ACF has no 
significant second peak.  
Assuming a 1$\sigma$ profile with width equal to the 1667 MHz 
line width, we obtain a rough upper bound on the 1665 MHz emission 
which provides a lower bound on the hyperfine ratio: $R_H\geq11.1$.
The DSS image shows a pair of overlapping galaxies with significant 
separation (roughly $11$\arcsec\ or 26 kpc).
%Weights smooth across OH spectrum.  RMS in area of expected 1665 MHz 
%line is 0.25 mJy.  
%separation 
%(12 16 14.9, 10 47 53 and 12 16 14.3, 10 47 46 => 0.6sec, 
%7\arcsec = 11.3\arcsec separation).  No HST.

\noindent{\bf 12549+2403: }
This OHM has a sharp, well-defined second peak in the ACF.  
The 1665 MHz line is poorly defined in the spectrum, and
if present, has a 3.6$\sigma$ peak.  There are mild standing waves 
in the bandpass which could masquerade as the 1665 MHz line.  
The 1667 MHz line is asymmetrical, with sharp red falloff and a blue tail.
The DSS image of this object is extended and elliptical in shape but 
more or less regular (no clear signs of interaction).   
%No HST.

\noindent{\bf 13218+0552: }
The host of this OHM is classified as a QSO or Seyfert 1 
(Low \etal 1988), and is one of three Seyfert 1 hosts of OH megamasers.
The other two are Mrk 231 ({\it IRAS} 12540+5708; Baan, Haschick, \& 
Henkel 1992; Baan, Salzer, \& LeWinter 1998; Kim, Veilleux, \& Sanders 1998)
and UGC 545 ({\it IRAS} 00509+1225; Bottinelli \etal 1990; Sanders \etal 1988),
and Tables \ref{OHM:opt} and \ref{OHM:rad} list the properties of these OHMs.
%It seems likely that Seyfert 1 AGN lack the molecular
%gas column required to produce OHMs.  This viewpoint can
%remain intact if the masing is produced in a second obscured nucleus or
%star formation region.
The OH spectrum of 13218+0552 shows two main broad emission peaks with
a separation of 490 km s\minusone\ in the rest frame which may be 
associated with multiple nuclei.  The FWHM listed in Table \ref{detectOH}
is measured from the strongest line.  The overall spectrum shows many 
significant peaks (at least 4) and is quite broad,
spanning 1560 km s\minusone in the rest frame at 10$\%$ of the peak
flux density.  Disentangling the 1665 and 1667 MHz lines is not possible
in this spectrum, so $W_{1667}$ is measured from the entire profile.  
Note that there are dips in the weights spectrum, two of which coincide
with the OH spectrum.  The small dip at 61100 km s\minusone\ does not 
correspond to any OH peak or feature, indicating that this RFI was properly
cleaned.  The broad feature in the weights spectrum at 62200 km s\minusone\ 
indicates the presence of a global positioning system downlink signal at 
1381 MHz.  This signal
could not be completely removed from the OH spectrum and was masked in the
final spectrum.  
Boyce \etal (1996) have imaged the host of this QSO with HST.  The host 
shows tidal features and possibly a double nucleus with separation 
less than 1 kpc.  Boyce \etal propose that while this object does not
technically qualify as an optical QSO (only as a Seyfert 1), it 
contains a luminous buried QSO\@.  
Lo, Chen, \& Ho (1999) obtain an upper bound on the H$_2$
mass of this object from CO(1--0) observations at the NRAO 12 m antenna:
$M(H_2) < 2.4 \times 10^{10} M_\odot$.
%It is also an X-ray source.  HST
%imaging available, indicating a disturbed morphology and perhaps a double 
%nucleus (max separation of 1 kpc).  ``Reddest known quasar''  ; buried QSO.
%F13218 has no second peak in the ACF (very smooth \& broad).  
%Weights spec shows a broad hole at 1381 MHz (GPS downlink?) and blips at 
%1374.5, 
%1379.0, 1385.0 of depth $< 5\%$, and quite narrow.  1385.0 falls into OH 
%spectrum,
%but does not correspond to any peak or feature (properly cleaned).  1665 
%extraction
%is futile.  Whole OH spec is measured.  GPS is masked.  Spectrum shows many 
%significant
%peaks (at least 4), and looks very much like F09531.  
%FWHM is measured for the strongest line (width at 10$\%$ level is 
%1.389188$-$1.381991=7.197MHz=1294km/s).  $W_{1667}$ is
%measured from the entire profile.  
%DSS image shows slight tidal features around an extended circular object 
%with a 
%single central bright point.  
%  HST archive:  WFPC2 (3x600s at F702W; 1900s at 300) \& FOS.  

\noindent{\bf 14043+0624: }  
This OHM has a well-defined but broad 1665 MHz line.  There is an absorption
feature at the edge of the 1665 MHz line which is not identified as RFI by
the weights spectrum.  If it is RFI which is stable in time or low-level, 
then it will not be identified and removed by the RFI flagging routine, and 
may also affect the 1665 MHz line.  This would explain the unusually broad
and strong 1665 MHz line which produces an anomalous hyperfine ratio below
the thermodynamic equilibrium value:  $R_H = 1.4$.  The depression in the
weights spectrum at 33450 km s\minusone\ marks locally generated RFI at
1500 MHz.  
%Weights is very clean (narrow RFI at 1500 MHz)...nothing at the abs feature.
%Slightly extended DSS image.   No HST.
%No spectral ident available.

\noindent{\bf 14059+2000: }  
The spectrum of this OHM has a few low-level features which 
may or may not be OH emission.  This spectrum has a strong main peak, 
a significant 1665 MHz line, and broad wings, especially on the red side.  
The width at $10\%$ of the peak flux density is surprising: 1650 km s\minusone\
in the rest frame.  The 1667 MHz line measurements 
exclude the red wing and the broad low-level blue emission feature, but all 
components are included in the total OH measurement.  \LOH\ is more than 
an order of magnitude larger in this OHM than $L_{OH}^{pred}$.
%1.487087$-$1.478952=8.135MHz=1463km/s (uncorrected for z).
%The weights spectrum is smooth across the OH lines.  
%Irregular DSS image, but barely resolved.  No HST.

\noindent{\bf 14070+0525: }  
This is an OH gigamaser, and the most distant OHM known at $z=0.2655$.
It was discovered by Baan \etal (1992), and redetected in this survey.
The OH lines fall nicely into a clean spectral window, and the spectrum
illustrates the difficulty of OH spectral line observations beyond 
$z\simeq0.2$.  The RFI has been truncated on the high velocity side of 
the spectrum.  The weights spectrum clearly
identifies the RFI seen in 
the spectrum, but the cleaning procedure does not remove all of the RFI
because it is present in every record.  Note that the weights spectrum
is clean across the OH spectrum.  The OH lines are broad and blended, and
identification of a 1665 MHz line is not possible.  The multiple broad 
peaks suggest multiple masing nuclei in this object.  The FWHM quoted in 
Table \ref{detectOH} is measured from the strongest component.  The 
width at 10$\%$ of peak flux density is 1580 km s\minusone\ in the rest frame.
A comparison of the spectrum in Figure \ref{spectra} and the spectrum in 
Baan \etal (1992) shows no significant spectral profile changes, although
the  flux calibration seems to be different between
the two:  Baan \etal  measure an integrated flux density of 1.01$\times10^4$
mJy km s\minusone, while we measure 6.11$\times10^3$ mJy km s\minusone.
The host of this OHM is classified a Seyfert 2 by Kim, Veilleux, \& 
Sanders (1998).
Solomon \etal (1997) measure a CO(1--0) line width of 270 km 
s\minusone, estimate a H$_2$ mass of 4.4$\times 10^{10} M_\odot$, 
and derive a blackbody radius of 284 pc.  
%ACF is extremely broad and shows no second peak. 
%DSS image is unresolved and quite faint.   No HST.

\noindent{\bf 14553+1245: }
This OHM has a well-defined second peak in the ACF which agrees with the 
1667 MHz prediction for the 1665 MHz line, but there is no clear spectral
feature.  We use the noise to set an upper bound on the 1665 MHz line and
a lower bound on the hyperfine ratio:    %RMS=0.19 mJy; 
using 1$\sigma$ and the width of 1667 line, $R_H \geq 14.5$.  The main OH 
line sits atop broad low-level emission which is included in the measure
of the total OH emission, while measurements of the 1667 MHz line use
only the most obvious feature.  
%Deep hole in weights at 1478.7 MHz,
%0.7 MHz wide at most (top).  Does not correspond to any significant spectral 
%feature (well-cleaned).  Dips at 1480.8, 1483.3 which do not correspond to 
%significant spectral features.  
%Peak at 1479.4 is tentatively identified as the 1665 MHz 
%line (3.1$\sigma$).  
%{\bf NOTE:} Peak tentatively ID'ed as 1665 is at INCORRECT frequency.  Measure 
%low-level stuff instead. -> exclude this line from total OH measure?
%Total OH includes the broad emission, while 1667 \& 1665 use just the
%obvious lines.  
%Unresolved DSS image.   No HST.

\noindent{\bf 14586+1432: }
The spectrum of this OHM is extremely broad and complicated.  
The FWHM listed in Table \ref{detectOH} is measured from the 
central (double) peak, but the FWHM of the
entire complex is 950 km s\minusone\ in the rest frame.
The total line complex spans 1160 km s\minusone\ (rest frame) at 
10$\%$ of peak flux density.  
The ACF is very smooth and shows no second peak due to the blending of
emission lines.  The weights spectrum indicates some narrow RFI on the 
edge of the OH profile which does not significantly affect the spectrum.
%Otherwise, if we go to the half power pt on the outside of the
%complex, FWHM=(1.455025-1.45042)=4.605 MHz=950 km/s.  
%ACF 2nd pk has little meaning in this 
%case (no pk anyway) --- ACF is v smooth.  One RFI region (narrow) in weights.
One interpretation of this complicated spectrum is straightforward:  there
are two nuclei in this object, both of which are masing.  We label nucleus A
the higher velocity nucleus, and nucleus B the lower velocity nucleus.  
Going from right 
to left in the OH spectrum, we see the peaks corresponding to 1665A, 1667A, 
1665B (on the shoulder of 1667A), and 1667B.  The predicted locations of the
of 1665 MHz lines from the positions of the 1667 MHz lines agrees nicely with
the actual emission peaks, and there is an uncanny correspondence of features
between 1667A and 1665A.  The optical redshift favors nucleus A, and we
use 1667A to measure $v_{1667}$ in Table \ref{detectOH}.  The two nuclei 
have a velocity difference of 516 km s\minusone\ in the rest frame.  
We use the entire complex to compute $W_{1667}$, which is an upper limit 
because it includes 1665 MHz emission.
%Think about changing the center for predicting 1665.  
%Do v1667.  1667b centroid = 1.45480753.  1667a centroid = 1.4523065.
%Put two prediction lines for 1665 on the plot --- both agree nicely with the 
%actual lines.  Preds come from the centroid of the two 1667 lines.  Correspondence of
%features in the two reddest lines is uncanny.  Need more time w/o sun.  v1667 
%is poorly defined here.  We used the centroid of the center line (1667a) for
%v1667.  
%W1667 is an upper limit, b/c it includes the 1665 fluxes 
%No HST.  DSS img barely or un resolved \& uninteresting.

\noindent {\bf 17161+2006:}
This OHM shows a strong main line and only weak evidence for a 1665 MHz line.
There is a weak second peak in ACF which corresponds with the predicted 
location of the 1665 MHz line from the center of the 1667 MHz line.  
The ACF was computed after masking the deep RFI trough at 31850 km s\minusone.
The weights spectrum is clean across the OH lines.  
The identification of the 1665 MHz line is tentative, and resembles 
other baseline features, making the hyperfine ratio uncertain.
%Small offset of optical and OH redshifts.  1665 line is not obvious 
%(hence the twiddle in $R_H$).  RFI misses OH lines.
%No HST.  
%DSS optical counterpart is not
%obvious --- there are several potential counterparts.  

\subsubsection{Notes on OH Absorbers}

\noindent{\bf 12107+3157: } 
This is an OH absorber, but unlike {\it IRAS} 19154+2704 (see Paper I), 
there is no
obvious 1665 MHz line, and no second peak in the ACF.  The absorption feature
is broad (420 km s\minusone\ in the rest frame), but the signal to noise
of this spectrum is low, so potentially interesting structure in the line 
will have to be confirmed by further observations.  The NVSS flux is quite high
for a ULIRG at $z=0.2065$ (32 mJy), indicating that there is a radio loud nucleus 
in this object.  

\subsection{Survey Completeness}\label{subsec:completeness}

\begin{figure}[!t]
%\centerline{\psfig{figure=Darling.fig3.epsi}}
\epsscale{1}
\plotone{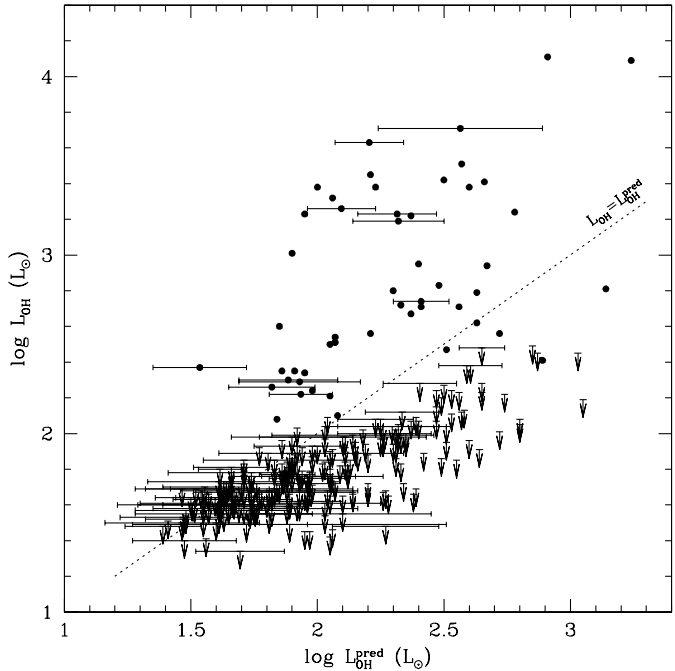}
\caption[Survey Completeness:  OH Luminosity Measurements versus Predictions]
{Survey completeness:  OH luminosity measurements versus predictions
for all unambiguous survey targets.  OH megamasers
are plotted as filled circles and nondetections are plotted arrows
to indicate upper limits on OH line luminosity.  Horizontal error
bars indicate objects with upper limits on 100\micron\ detection;
the bars span the available range of \LFIR\ set by the upper 
detection limit on $f_{100\mu m}$ and negligible 100\micron\ flux.  
The heavy dashed line indicates the locus $L_{OH} = L_{OH}^{pred}$.
\label{completeness}}
\end{figure}

The Arecibo OH megamaser survey is flux-limited and thus will not identify
OHMs which fall below the detection threshold.  How many OHMs is
the survey missing?  In order to predict the OH luminosity of OHM candidates
in the survey, we employed the relation obtained by Kandalian (1996) from
the then known OHM sample:  
$\log L_{OH}^{pred} = (1.38\pm0.14) \log L_{FIR} - (14.02\pm 1.66)$. 
Rather than predicting \LOH\ of OHMs, the predicted OH luminosity seems to 
indicate a rough division between OHMs and nondetections, as shown in 
Figure \ref{completeness} (since $\log L_{OH}^{pred} \propto \log L_{FIR}$, 
this figure is identical to Figure 
\ref{survey_relation} --- only the scale on the abscissa has changed).
There is little overlap between the OHMs and the nondetections, indicating
a well-defined line detection threshold, 
and there are likely to be OHMs remaining in the nondetections 
with $L_{OH} < 10^2 L_\odot$.  Taking into account the OHM fraction as a 
function of \LFIR\, (see \S \ref{sec:ohfrac}), the OH-FIR relation
determined in \S \ref{sec:relation}, and the scatter in the relation,
we estimate that there are roughly a dozen undetected OHMs lurking
in the nondetections, nearly all of which have $L_{OH} < 10^2 L_\odot$.

\section{The Hosts of OH Megamasers\label{sec:hosts}}

The complete Arecibo OH megamaser survey has well-defined selection
criteria and an adequate sample size for an investigation of the relationships
between the flux-limited sample of OHMs and the merging systems which 
produce them.  We investigate the OH megamaser fraction in (U)LIRGs
as a function of their properties, the nature of the star formation
in OHM hosts, and the relationships between the properties of OHMs and 
their hosts.  

\subsection{The Received View}\label{sec:known}

All known OHMs have been observed in
luminous infrared galaxies (LIRGs), strongly favoring the most FIR-luminous, 
the ultraluminous infrared galaxies (ULIRGs; Baan 1991).  
Photometric surveys have shown the ULIRGs to be 
nearly exclusively the product of galaxy mergers (Clements \etal 1996).  
VLBI measurements have shown 
that OHMs are ensembles of many masing regions which originate 
in the nuclear regions of (U)LIRGs on scales 
of a few hundred parsecs or less (Diamond \etal 1999).  
OHM activity requires: (1) high molecular density, (2) a pump
to invert the hyperfine population of the OH ground state, and (3) a
source of 18 cm continuum emission to stimulate maser emission 
(Burdyuzha \& Komberg 1990).
The environments produced in merging galaxies can supply all of 
these requirements:  the merger
interaction concentrates molecular gas in the merger nuclei, 
creates strong FIR dust emission from reprocessed starburst
light and AGN activity, and produces radio continuum emission from AGN
and starbursts. The FIR radiation field can invert the OH population
via the pumping lines at 35 and 53 $\mu$m (\cite{ski97}).  
Masing can then be stimulated 
by 18 cm continuum emission from starbursts or AGN, or even by spontaneous 
emission from the masing cloud itself (Henkel, G\"{u}sten, \& Baan 1987).

Based on optical morphology and surface brightness profiles,
the FIR luminosity of LIRGs seems to be correlated with the stage in a
merger sequence such that late-stage mergers are the most FIR-luminous 
(Sanders, Surace, \& Ishida 1999).  The OHM fraction in LIRGs is a strong 
function of \LFIR\ (Baan 1991), which indicates that the later stages of
merging may be preferred for OHM formation.  
This makes some physical sense, based on the high molecular gas 
density required to produce OHMs ($n_{H_2}=10^{4-7}$ cm$^{-3}$; 
Baan 1991).  Early in the merger sequence, infall and concentration
of molecular gas in the nuclear regions is just beginning, whereas late
in the merger sequence, clouds are disrupted by ionizing radiation, 
a superwind phase, or a QSO eruption.  
If OHMs mark a specific phase in major mergers, then they provide
useful tracers of the galaxy merger rate as a function of redshift, 
particularly since they may be observed at cosmological distances with
current instrumentation (Baan 1989; Burdyuzha \& Komberg 1990; Briggs 1998)

\subsection{The Known OH Megamasers}

The library of OHMs is incomplete and drawn from many disparate surveys,
and a comprehensive study strictly from the literature is difficult.
We present here a compilation of the known OHMs which is useful
for discussing the received view about OHM environments and mechanisms, and
which for the first time attempts to gather all of the available 
information into a single place and format.  
Baan, Salzer, \& LeWinter (1998) present the largest compilation of OHMs 
available in the literature (50 objects).  This includes 
3 OHMs which do not appear anywhere
else in the literature.  To this list of 50 OHMs, we add 4 OHMs found in 
the literature.  
Of these 54 objects, 25 OHMs do not have published spectra, and 5 
are highly suspect detections.  Six of the OHMs listed are not technically 
OHMs ($L_{OH} = 10^1$--$10^4 L_\odot$) or OH ``gigamasers'' 
($L_{OH} > 10^4 L_\odot$); they are OH ``kilomasers.''  Tables \ref{OHM:opt}
and \ref{OHM:rad} summarize the optical redshifts and FIR properties and the OH
and 1.4 GHz properties of the known OHMs (in the loose sense).  

We make every effort to present uniform OHM data, which is made 
difficult by sparse and sometimes conflicting reporting of data.
When measured properties conflict, we preferentially use the 
observations presented in detection papers.  The peak
flux density of the 1667 MHz OH line is generally not quoted in the literature.
Data on the hyperfine ratio, the OH velocity, and the OH line width 
are most often lacking.
In order to place the data into a uniform cosmology, we take observer units
of integrated line flux, usually erg s\minusone\
cm$^{-2}$, and convert to a cosmology-dependent line luminosity.

Tables \ref{OHM:opt} and \ref{OHM:rad} list respectively 
the optical/FIR and radio properties of the 54 known OH 
megamasers and kilomasers (hereafter jointly referred to as OHMs).
Table \ref{OHM:opt} lists the optical redshifts
and FIR properties of the OHMs in a format identical to Table 
\ref{nondetectFIR}.  
Table \ref{OHM:rad} lists the OH luminosity, 1.4 GHz 
flux density, and nuclear classification
of the known OHMs in the following format:
Column (1) lists the {\it IRAS} FSC name, as in Table \ref{OHM:opt}.
Column (2) lists the heliocentric optical redshift, as in Table \ref{OHM:opt}.
Column (3) lists the $\log$ \LFIR, as in Table \ref{OHM:opt}.
Column (4) lists the predicted isotropic OH line luminosity, 
$\log L_{OH}^{pred}$,
based on the Malmquist bias-corrected $L_{OH}$-\LFIR\ relation 
determined by Kandalian (1996) from 49 OHMs:  
$\log L_{OH} = (1.38\pm0.14) \log L_{FIR} - (14.02\pm 1.66)$ 
(see \S \ref{sec:relation}).
Column (5) lists the logarithm of the measured isotropic OH line luminosity, 
which includes the 
integrated flux density of both the 1667.359 and the 1665.4018 MHz lines.
Note that $L_{OH}^{pred}$ is generally 
less than the actual \LOH\ detected (42 out of the 51 objects with 
available OH measurements).
Column (6) lists the peak flux density of the 1667 MHz OH line in mJy.
Column (7) lists references for the listed OH line properties.
Column (8) lists the 1.4 GHz continuum fluxes, from the NRAO 
VLA Sky Survey (NVSS; Condon et al. 1998).  If no continuum source lies within 
30\arcsec\ of the {\it IRAS} coordinates, an upper limit of 5.0 mJy is listed. 
Note that sources south of $-40^\circ$ declination are not included in 
the NVSS.
Column (9) lists the optical spectroscopic classification, if available:
(S1) Seyfert 1,
(S1.5) Seyfert 1.5,
(S2) Seyfert 2,
(A) active nucleus,
(C) composite active and starburst nucleus,
(H) \HII region (starburst), and 
(L) low-ionization emission region (LINER).  
References for the classifications are shown in parentheses and 
included at the foot of the table.
Column (10) lists source notes, which are given at the foot of the table.

\begin{figure}[t]
%\centerline{\psfig{figure=Darling.fig5.epsi,clip=t,width=6in}}
\epsscale{1}
\plotone{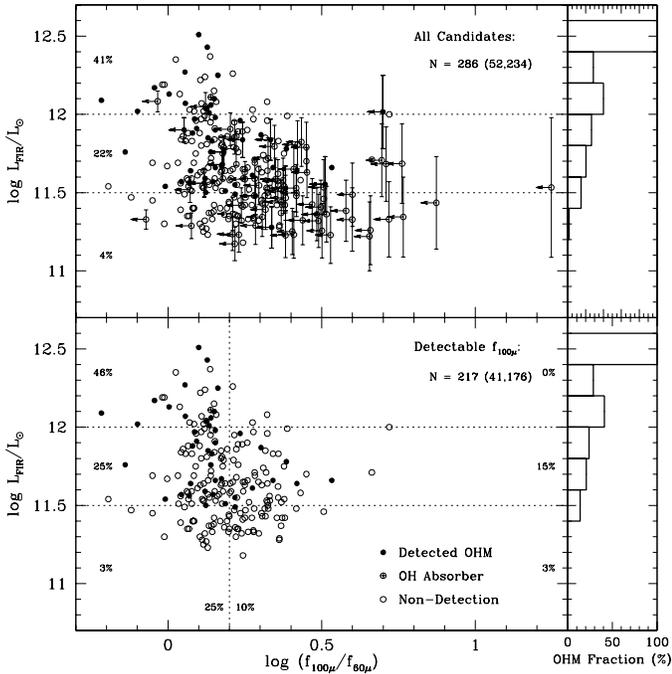}
\caption[The Arecibo OH Megamaser Survey FIR Color-Luminosity Plots]
{The Arecibo OH megamaser survey FIR color-luminosity plots.
The two left panels show \LFIR\
versus FIR color for candidates observed to date, and the two right panels
show the OHM fraction as a function of $L_{FIR}$.  Filled circles mark OHMs, 
empty circles mark non-detections, and the crossed circle marks the OH
absorber.  Points with error bars are non-detections at 100 $\mu$m.  
Vertical error bars indicate the possible range of $L_{FIR}$, constrained
by $f_{60\mu m}$ and an upper limit on $f_{100\mu m}$.  Horizontal arrows
indicate upper limits on FIR color.  Inset percentages indicate the OHM
fraction for each sector delineated by the dashed lines.  The upper panels
plot all 286 candidates observed.  The lower panels plot the 217 objects
with detected $f_{100\mu m}$.  The inset numbers follow the key:  
N $= $ Observed (OHMs, Non-Detections).  
\label{survey_color}}
\end{figure}

\subsection{The FIR Luminosities and Colors of OH Megamaser Hosts}
\label{sec:ohfrac}

The Arecibo OHM survey detected
1 OHM in every 5.5 candidates, but the OHM fraction is not constant across all
LIRGs as shown in Figure \ref{survey_color}.  The OHM fraction 
is a strong function of \LFIR, increasing to at least one in three for ULIRGs.
A similar trend was quantified by Baan (1991).
Clearly, some property of the most luminous LIRGs promotes OHM production.  
Note also that nearly half of the LIRGs in the upper left quadrant of the 
lower plot in Figure \ref{survey_color} host OHMs, indicating a color 
dependence on OHM production as well.  This color dependence was also 
noted by Unger \etal (1986), Staveley-Smith \etal (1992), 
Baan, Haschick, \& Henkel (1992), and others.  

There are indications in Figure \ref{survey_color} that LIRGs with ``warmer'' 
FIR colors (smaller $f_{100\mu m}/f_{60\mu m}$) are more likely
to host OHMs.  The top panel of Figure \ref{survey_color}, which 
plots all unambiguous objects in the survey, shows that a quarter of the
objects are undetected by {\it IRAS} in the 100 \micron\ band.  A proper
analysis of the heavily censored FIR color of OHM hosts requires 
survival analysis techniques which can properly account for upper limits
on measured quantities.  These 
techniques are available in the IRAF ASURV package  
Rev 1.2 (LaValley, Isobe, \& Feigelson 1992), which implements the 
methods presented in Feigelson \& Nelson (1985).
The tests employed by ASURV find a difference between the colors of
OHM hosts and nondetections at the 0.5--1.4$\%$ significance level, and
the Kaplan-Meier estimates of the mean colors 
$\log (f_{100\mu m}/f_{60\mu m})$ 
are $0.12\pm0.02$ for the OHM hosts and $0.18\pm0.01$ 
for the nondetections.  Hence, despite the obfuscation of the upper
limits on colors, the apparent preference for warm LIRGs to produce
OHMs is confirmed with high confidence.

\begin{figure}[!t]
%\vspace{-10pt}
\epsscale{1}
\plotone{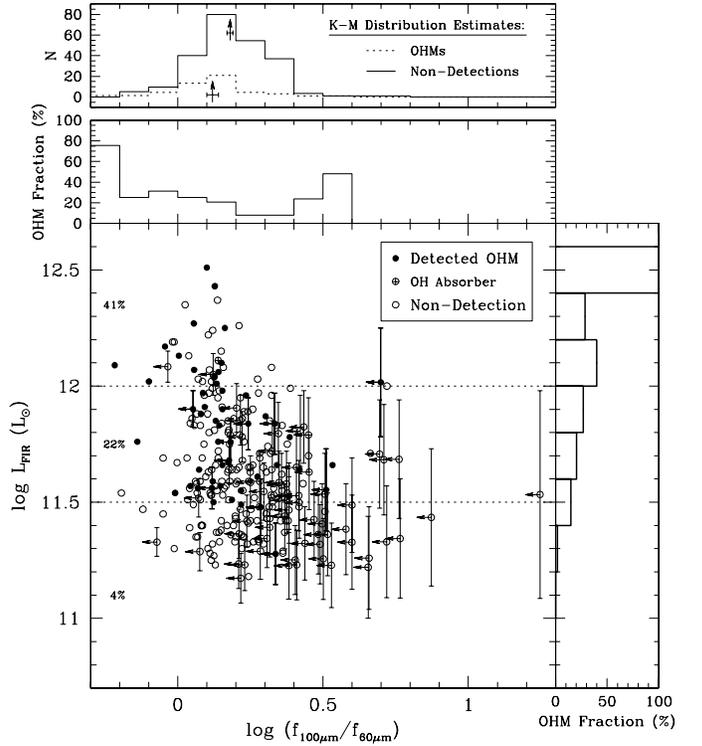}
%\centerline{\psfig{figure=Darling.fig6.epsi,clip=t,width=6in}}
%\vspace{-5pt}
\caption[The OH Megamaser Fraction by FIR Luminosity and Color]
{The OH megamaser fraction by FIR luminosity and color.
The main panel shows \LFIR\
versus FIR color for unambiguous OHM candidates, the right panel
shows the OHM fraction as a function of $L_{FIR}$, the topmost panel
shows the color distribution of OHMs and nondetections estimated
by a survival analysis, and the middle panel shows the estimated OHM
fraction as a function of FIR color.  
Points with error bars are non-detections at 100 $\mu$m.  
Vertical error bars indicate the possible range of $L_{FIR}$, constrained
by $f_{60\mu m}$ and an upper limit on $f_{100\mu m}$.  Horizontal arrows
indicate upper limits on FIR color.  Inset percentages indicate the OHM
fraction for each sector delineated by the dashed lines.  The arrows in the
top panel indicate the mean and standard deviation survival analysis 
estimates for the OHM and nondetection populations.  
\label{color:survival}}
\end{figure}

Figure \ref{color:survival}
shows the FIR color-luminosity plot of all unambiguous survey objects with
the OHM fraction as a function of both luminosity and color.  The 
fraction versus color is derived from the Kaplan-Meier estimates of the
differential distribution of the OHMs and the nondetections.  These 
distributions are plotted in the top panel of Figure \ref{color:survival}.
Note that the OHM fraction, while large at the highest and lowest color 
values, 
is derived from very few objects.  The reliable and noteworthy information 
contained in this analysis is the strong rise in the fraction of OHMs 
with warmer color from $\log (f_{100\mu m}/f_{60\mu m})$ values of 
$\sim$0.4 to $\sim$0.  This is also apparent in the differing shapes of
the estimated distributions of OHMs versus nondetections.
We confirm that LIRGs with warmer FIR colors --- and hence higher dust
temperatures --- are more likely to produce OHMs.  
ULIRGs generally have higher dust temperatures than the less luminous
LIRGs, so there may be some entanglement of the roles of \LFIR\ and 
FIR color in producing OHMs.  It is unclear if the two effects can be
disentangled.  Higher dust temperatures and higher luminosities will both
increase the number of photons available to pump OH molecules, so they
may both be related to \LOH .

\begin{figure}[t!]
%\centerline{\psfig{figure=Darling.fig7.epsi}}
\epsscale{1}
\plotone{Darling.fig6.epsi}
\caption[The Arecibo OH Megamaser Survey:  \LOH\ versus \LFIR]
{The Arecibo OH Megamaser Survey:  \LOH\ versus \LFIR\ for all 
unambiguous survey targets.  OH megamasers
are plotted as filled circles and nondetections are plotted arrows
to indicate upper limits on OH line luminosity.  Horizontal error
bars indicate objects with upper limits on 100\micron\ detection;
the bars span the available range of \LFIR\ set by the upper 
detection limit on $f_{100\mu m}$ and negligible 100\micron\ flux.  
\label{survey_relation}}
\end{figure}

\begin{figure*}[!ht]
%\centerline{\psfig{figure=Darling.fig8.epsi,clip=t,width=6in}}
\epsscale{1.7}
\plotone{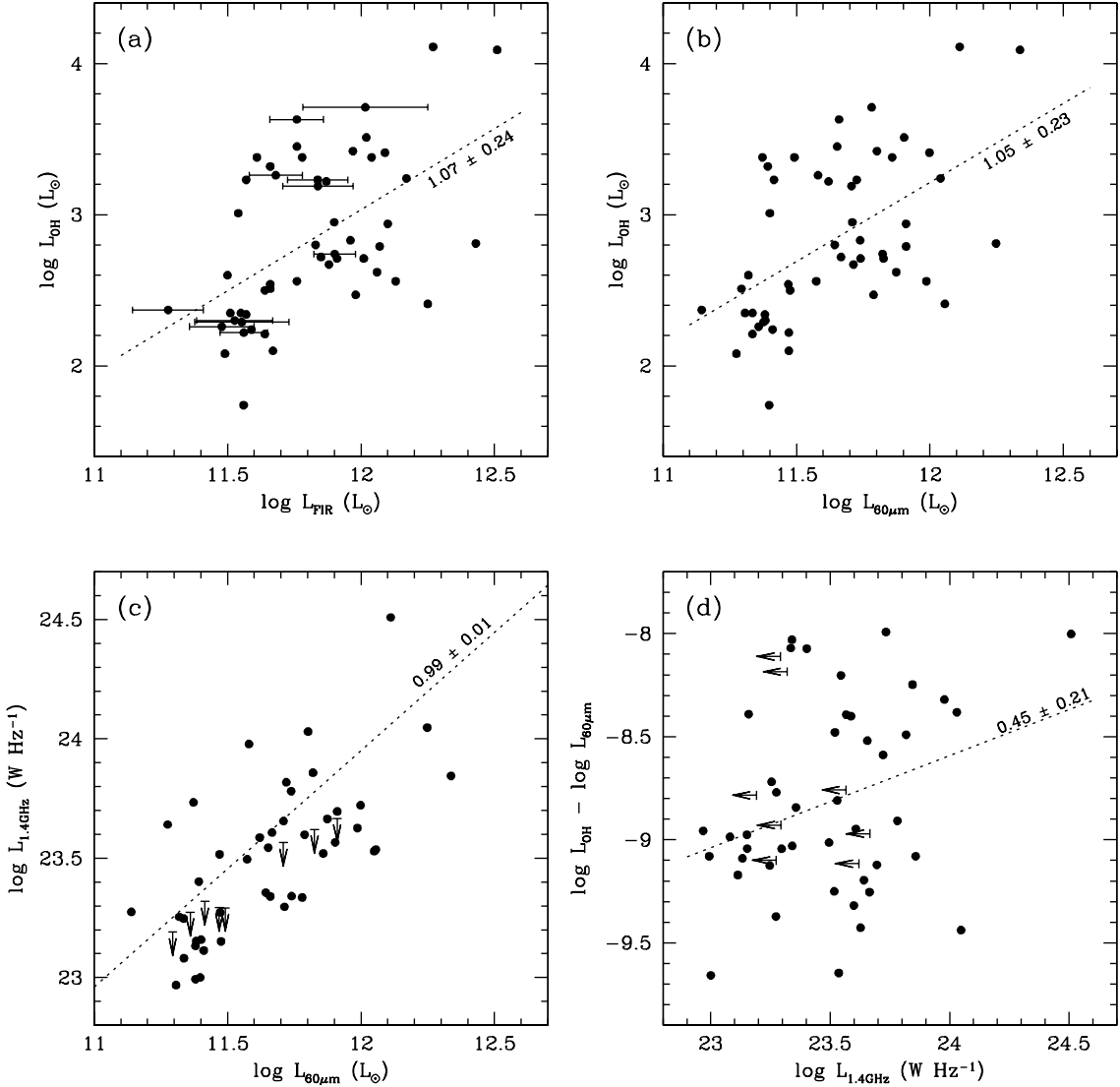}
\caption[OH-FIR-Radio Relations of Arecibo Survey OH Megamasers]
{OH-FIR-Radio relations of Arecibo survey OH megamasers.  Upper limits
are indicated by arrows, and the available ranges for $L_{FIR}$ are 
indicated by error bars.  
Dotted lines are fits to the data labeled by their slopes except for
the $L_{1.4GHz}$ versus $L_{60\mu m}$ plot which is {\it not} at fit
to the data but instead 
shows the radio-FIR relation obtained from a large sample of
galaxies spanning many decades in luminosity (Yun, Reddy \& Condon 2001).
Shown are 
{\bf (a)} the OH-FIR relation, 
{\bf (b)} the OH-60\micron\ relation, 
{\bf (c)} the radio-60\micron\ relation, and
{\bf (d)} the unbiased OH-60\micron-radio relation.
\label{relations:ao}}
\end{figure*}

\subsection{The FIR-OH Relationship}\label{sec:relation}

Suppose that an OH maser is radiatively pumped by the FIR radiation field.
In a simple scenario of low gain unsaturated masing, the maser 
output is proportional to the pumping rate and to the stimulated 
emission rate (Baan 1989).  If the 
pump is the FIR radiation field, and the source
of stimulated emission is the radio continuum field, which is itself
proportional to the FIR radiation field in star-forming regions (Yun, Reddy, 
\& Condon 2001),
then the observed maser output can be related to a single parameter:
$L_{OH}\propto L_{FIR}\ L_{1.6GHz}\propto L_{FIR}^2$.
For the case of low gain saturated masing, the stimulant cannot be 
fully accommodated by the inverted gas, and drops out of the relation:
$L_{OH}\propto L_{FIR}$.
If we suppose that an OH megamaser represents an ensemble of many 
individual masing regions with different saturation states, then it
is reasonable to expect that $L_{OH} \propto L_{FIR}^\gamma$, 
where $1 < \gamma< 2$.  The measurement of $\gamma$ has
traditionally been frustrated by small samples, survey biases, and
theoretical predisposition.  However,  
Kandalian (1996) demonstrated that a simple regression of the available 
OHM data indicates $\gamma=1.66$, and when one properly accounts for 
Malmquist bias (\ie\ the 
correlation of \LOH\ and \LFIR\ with distance in flux-limited surveys),
the relationship becomes more linear:  $\gamma = 1.38 \pm 0.14$.  
VLBI observations of a few individual OHMs by Diamond \etal (1999) 
support the statistical results.  They find that OHM emission is likely
to be segregated into two emission regimes:  unsaturated extended
emission and high gain saturated compact ($<1$ pc) masing regions.  As seems
reasonable, OHMs appear to represent an aggregate of a broad
range of masing conditions.  
%This begs the question:  Why don't we
%observe only one type of masing in some galaxies, since there are many
%which show no evidence of masing whatsoever?  Do we observe such systems?
%Would we know it if we did? ***

Although the assumption of low gain masing in all OHMs and the use
of $\gamma$ as a saturation index ($\gamma=(2)1$ for (un)saturated masing)
is not supported by observations, we examine the empirical relationships
between $L_{OH}$, $L_{FIR}$, $L_{60\mu m}$, and $L_{1.4GHz}$ to 
identify and quantify trends in the OHM sample.  
Recall the basic relationship for masing under scrutiny:
\begin{eqnarray}
	L_{OH}  & \propto & L_{FIR}\ L_{1.6GHz}^{\gamma-1}.\label{eqn:relation}
\end{eqnarray}
This assumes radiative
pumping.  A good proxy for $L_{1.6GHz}$ is $L_{1.4GHz}$, a proxy for 
$L_{FIR}$ is $L_{60\mu m}$, and the 
radio continuum-FIR relationship for star forming galaxies provides
a nearly linear relationship between $L_{1.4GHz}$ and $L_{60\mu m}$
(Yun, Reddy, \& Condon 2001; see \S \ref{sec:fir-radio}).
Hence, we make several different determinations of $\gamma$ from fits
to \LOH\ versus \LFIR, \LOH\ versus $L_{60\mu m}$, and 
$L_{OH}/L_{60\mu m}$ versus $L_{1.4GHz}$.  These fits require progressively
fewer assumptions about the validity of the relationships between 
luminosities.  We determine the OH-FIR
relation of OHMs first from the Arecibo survey sample, then from the 
sample of all available OHMs.

\begin{figure*}[ht]
\epsscale{1.7}
\plotone{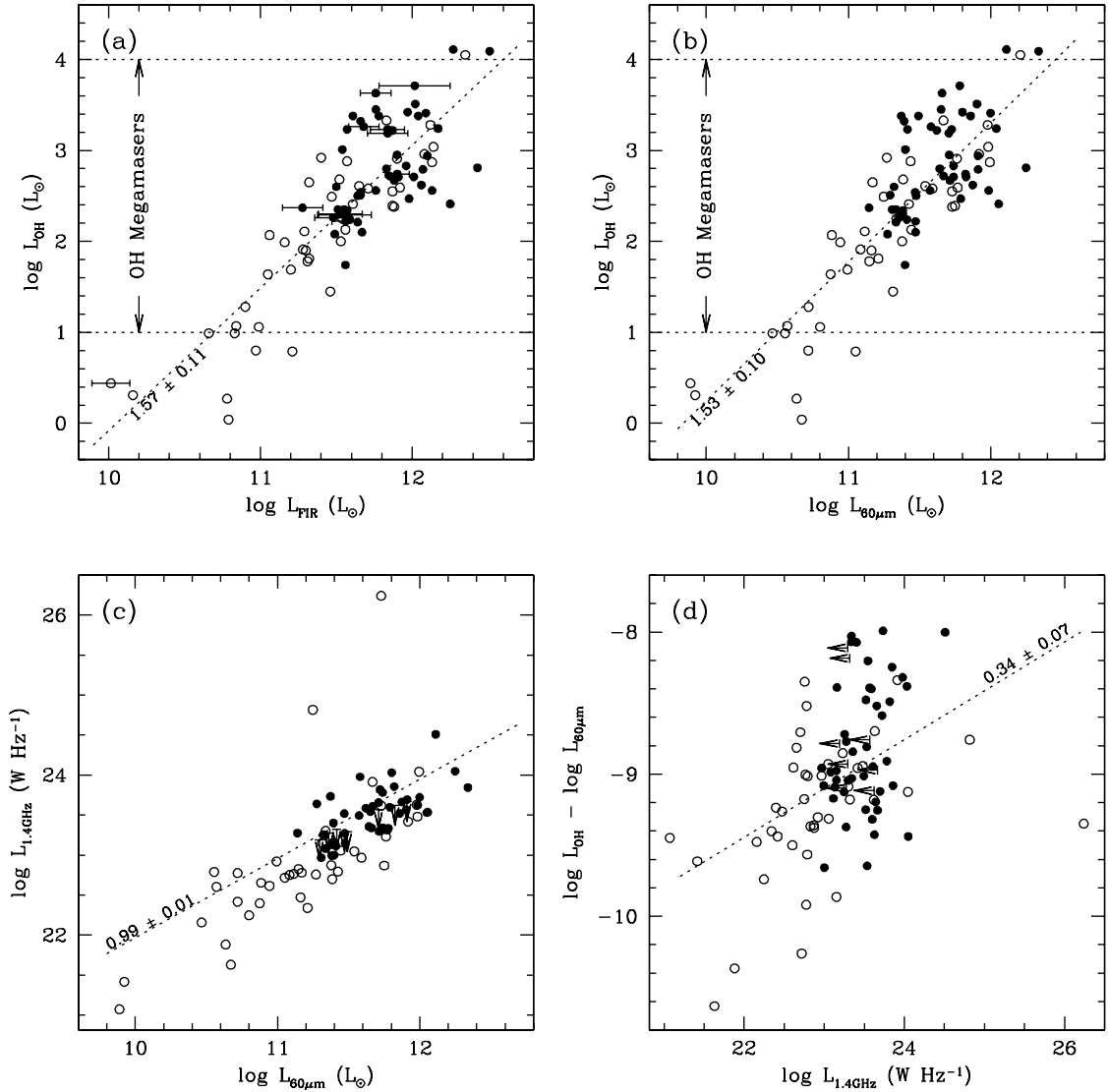}
%\centerline{\psfig{figure=Darling.fig9.epsi,clip=t,width=6in}}
\caption[OH-FIR-Radio Relations of All OH Megamasers]
{OH-FIR-Radio relations of all OH megamasers.  Filled circles indicate
Arecibo survey detections (including 3 previously discovered OHMs) and
open circles mark previously known OHMs.  
Upper limits
are indicated by arrows, and the available ranges for $L_{FIR}$ are 
indicated by error bars.  
Dotted lines are fits to the data labeled by their slopes except for
the $L_{1.4GHz}$ versus $L_{60\mu m}$ plot which is {\it not} at fit
to the data but instead 
shows the radio-FIR relation obtained from a large sample of
galaxies spanning many decades in luminosity (Yun, Reddy \& Condon 2001).
Shown are 
{\bf (a)} the OH-FIR relation, 
{\bf (b)} the OH-60\micron\ relation, 
{\bf (c)} the radio-60\micron\ relation, and
{\bf (d)} the unbiased OH-60\micron-radio relation.
\label{relations:both}}
\end{figure*}

\subsubsection{Arecibo Survey OHMs}
This sample includes all 52 OHMs detected in the Arecibo OHM survey, including
3 redetections ({\it IRAS} 10378+1108, 12018+1941, and 14070+0525).  The 
Arecibo survey is flux-limited, which indicates that there may be undetected 
OHMs still lurking in the sample.  Figure \ref{survey_relation} shows the
OH-FIR relation for all unambiguous survey targets, plotting OH nondetections
as upper limits on \LOH.  There is little overlap between the 
body of OHMs and nondetections in this plot, indicating that the detection
threshold is well-defined.  Note that a flux limit in a 
spectral line survey does not correspond cleanly to a luminosity limit at
a fixed distance when the variation of line profiles is large.

Figure \ref{relations:ao} plots the three permutations of the OH-FIR
relationship for the Arecibo OHM sample with fits indicated by dotted
lines and labeled by their slopes.  Also plotted is the radio
continuum-FIR relation for the sample with a dotted line indicating the
relationship derived by Yun, Reddy, \& Condon (2001).  
{\it This line is not a fit}.  Upper limits on the radio continuum
luminosity derived from the NVSS are indicated by arrows.
The data for these plots can be found in Tables \ref{detectFIR} and
\ref{detectOH} and the equivalent tables in Paper I and Paper II.

{\bf OH-FIR:}  
\LOH\ and \LFIR\ are not well-correlated ($R=0.53$), despite their mutual
correlation with distance (see Figure \ref{relations:ao}).  
Also, \LOH\ is not as correlated with 
distance as is \LFIR:  $R(L_{OH},D_L) = 0.57$; $R(L_{FIR},D_L) = 0.79$.
A simple fit finds the relation 
%\begin{equation}
$	\log L_{OH} = (1.07\pm0.24)\log L_{FIR} -(9.81\pm2.85). $
%\end{equation}
However, since both \LOH\ and \LFIR\ are correlated with luminosity
distance $D_L$, the partial correlation coefficient\footnote{The 
partial correlation 
coefficient between $x$ and $y$ at fixed $z$ is 
%\begin{displaymath}
$	R_{x\,y,\,z} = {R_{x\,y} - R_{x\,z} R_{y\,z}\over
			\sqrt{(1-R_{x\,z}^2)(1-R_{y\,z}^2)}}$
%\end{displaymath}
where $R_{i j}$ is the standard correlation coefficient.}
of \LOH\ with \LFIR\ at fixed $D_L$ further
reduces the OH-FIR correlation to non-significance:  $R=0.16$. %$S_r=1.12$
The regression corrected for Malmquist bias obtains $\gamma=0.32$.
Note that the correlation of \LOH\ with \LFIR\ at fixed $D_L$ is not
equivalent to the correlation between OH and FIR fluxes because each 
object lies at a different distance.  
%$m=0.32$, $b=-0.98$.

{\bf OH-60\boldmath$\mu$m:}  Analysis of the relationship between \LOH\ and 
$L_{60\mu m}$ produces nearly identical results to the \LOH-\LFIR\
analysis, despite the notion that the 60 \micron\ luminosity is a more
relevant diagnostic of the radiative IR pumping lines of the OH molecule
at 35 and 53 \micron.  Again, 
\LOH\ and $L_{60\mu m}$ are not well-correlated ($R=0.54$), despite the mutual
correlation with distance, and  \LOH\ is not as correlated with 
distance as is \LFIR: $R(L_{OH},D_L) = 0.57$; $R(L_{FIR},D_L) = 0.80$.
The uncorrected fit produces the relation 
$	\log L_{OH} = (1.05\pm0.23)\log L_{60\mu m} -(9.35\pm2.69)$.  
The corrected correlation is not significant ($R=0.16$)
and the corrected slope is $\gamma = 0.32$.
% $S_r=1.15$, $m=0.32$, $b=-0.87$.

{\bf OH-60\boldmath$\mu$m-Radio:}  
Since the OHMs in the sample might not exactly follow the radio-FIR 
relationship, assuming \LFIR$\propto L_{1.4GHz}$ may introduce 
extra scatter into the OH-FIR relation.  Instead --- still assuming that
the maser is radiatively pumped --- we investigate the dependence on 
the radio emission, which is directly related to the saturation state
of the maser.  Rearranging Equation \ref{eqn:relation}, and using
the 1.4 GHz radio continuum luminosity as a good proxy for the stimulating
radiation, we obtain a relationship which is not subject to Malmquist
bias and which incorporates more information about the masing process:
\begin{equation}
	\log L_{OH} - \log L_{60\mu m} = (\gamma-1)\log L_{1.4GHz} + \beta .
\end{equation}
All units are absorbed into the $\beta$ term.  
Requiring radio continuum detections in the NVSS reduces the sample to 
41 new OHMs
and 3 redetections (44 total).   Upper limits on radio luminosity are shown 
as arrows in Figure \ref{relations:ao}.  Fits do not incorporate OHMs with
upper limits on radio luminosity.
The correlation between $\log L_{OH}/L_{FIR}$ and $\log L_{1.4GHz}$ is 
poor ($R=0.31$) and the measured slope yields $\gamma=1.45\pm0.21$.
%For y vs radio:  $a=0.45\pm0.21$, $b=-19.31\pm4.91$.\\ 

%A linear regression 
%finds $\gamma = 1.27 \pm 0.08$ which is indistinguishable from the
%corrected value measured from the \LOH-$L_{60\mu m}$ relation, 
%but with larger uncertainty.  Figure \ref{relations:known}d shows that 
%the correlation is poor in this case, with
%correlation coefficient of 0.45 (compared to the unbiased coefficient 
%found from \LOH\ versus $L_{60\mu m}$ of 0.79).  The scatter in the 
%radio component of OHMs does provide some insight into OHM mechanism:  
%note that OHM activity seems to be uninfluenced by ``monster'' 
%radio activity.  This likely precludes the formation of OHMs associated
%with jets. ***

%***Finally, one should reconsider the assumption of radiative pumping in OHMs.
%If collisional pumping is responsible for a portion of OHM emission, then
%there will be scatter in the OH-FIR relation which cannot be accounted
%for by measurement errors.  It is possible that a tell-tale of collisional 
%pumping would be narrow, strong OH lines, so one might expect OH width
%to correlate with \LOH/\LFIR.  Such a correlation was found by Staveley-Smith
%\etal (1992) in a sample of 14 OHMs.  ***

\subsubsection{Arecibo Survey + Known OHMs}
Redetections of previously known OHMs are given the OH values measured 
in the survey.  We do not wish to take credit away from those
who discovered these objects, but do favor including the higher 
signal to noise measurements of the Arecibo OHM survey.  Using
the values measured in the survey also enhances the uniformity
of the sample for these analyses.  
There are 43 previously known OHMs, 49 new OHMs, and 3 redetections 
(95 data points) in this expanded sample.  
Figure \ref{relations:both} reproduces Figure \ref{relations:ao}
for the expanded sample with the two OHM subsets as labeled.
The data for the previously identified OHMs can be found in 
Tables \ref{OHM:opt} and \ref{OHM:rad}.

{\bf OH-FIR:}  
\LOH\ and \LFIR\ are well-correlated ($R=0.84$), and are correlated with
distance:  $R(L_{OH},D_L)=0.71$; $R(L_{FIR},D_L)=0.74$.  
A fit obtains the relation 
$	\log L_{OH} = (1.57\pm0.11)\log L_{FIR} -(15.76\pm1.22). $
Unlike the Arecibo OHM sample alone, the combined sample retains 
a significant OH-FIR correlation after Malmquist bias correction: $R=0.66$. 
The corrected slope in the relation is shallower than the simple fit slope, 
at $\gamma = 1.24$.
%\indent	$R=0.84$; 
%OH-FIR:  $a = 1.57\pm0.11$, $b = -15.76\pm1.22$. \\
%\indent Corrected relation:  $R=0.66$, $S_r=7.63$, $m=1.24$, $b=-11.96$.

{\bf OH-60\boldmath$\mu$m:}  
Analysis of the relationship between \LOH\ and $L_{60\mu m}$ 
again produces nearly identical results to the \LOH-\LFIR\
analysis.  \LOH\ and $L_{60\mu m}$ are correlated with each other
and with distance:  $R=0.84$; $R(L_{OH},D_L)=0.71$; $R(L_{60\mu m},D_L)=0.72$.
The uncorrected fit obtains the relation 
$	\log L_{OH} = (1.53\pm0.10)\log L_{60\mu m} -(15.04\pm1.19)$,
and the corrected fit retains correlation between \LOH\ and 
$L_{60\mu m}$ ($R=0.66$) and has a slope of $\gamma=1.21$.
%$a = 1.53\pm0.10$, $b = -15.04\pm1.19$.  \\
%\indent Corrected relation:  $r=0.66$, $S_r=7.62$, $m=1.21$, $b=-11.42$.

{\bf OH-60\boldmath$\mu$m-Radio:}  
Requiring radio continuum detections in the NVSS reduces the sample
to 41 previously known OHMs, 41 new OHMs, and 3 redetections (85 total).   
Upper limits on the radio continuum luminosity are shown 
as arrows in Figure \ref{relations:both}.  Fits do not 
incorporate OHMs with upper limits on the radio luminosity.
The correlation between $\log L_{OH}/L_{FIR}$ and $\log L_{1.4GHz}$ is 
again poor ($R=0.46$) and the measured slope yields $\gamma=1.34\pm0.07$.
%OH-60\micron-radio:  %$a=0.34\pm0.07$, $b=-11.42\pm1.68$.\\ <= ???
%Correlation is poor:  $r=0.46$. \\
%For radio vs y:  $a=0.62\pm0.13$, $b=28.85\pm1.19$.\\
%For y vs radio:  $a=0.34\pm0.07$, $b=-16.99\pm1.68$.\\

\subsubsection{Discussion}  
As indicated by the poor correlation between variables in the 
fits of Arecibo survey OHMs, the intrinsic scatter in the OH-FIR relation 
is similar to the span of the survey in \LFIR\ and \LOH.  Because 
the survey work was strictly above $z=0.1$, only the upper end of 
the LIRG population was sampled, which does not provide an adequate
lever arm to extract a useful relationship from the intrinsic scatter
in the OH-FIR relationship.  Expanding the sample to all known OHMs
provides the span in \LFIR\ and \LOH\ required to resolve the 
relationship.  

The relationship derived from the expanded sample reveals a similar
slope for all three permutations of the OH-FIR relation. We adopt a 
final form for the Malmquist bias-corrected relation which applies
to both \LFIR\ and $L_{60\mu m}$:
\begin{equation}
 	\log L_{OH} = (1.2\pm0.1) \log L_{FIR} - (11.7\pm1.2).
\end{equation}
The slope of this relation is even shallower than the value determined 
from the previously known OHM sample by Kandalian (1996).  
The value of $\gamma=1.2$ suggests
that either most OHMs are nearly saturated or that global properties
of mergers such as \LFIR\ are not completely relevant to the production of
OHMs on small scales.  The latter conclusion is supported by the large 
scatter in the OH-FIR relationship.  The dependence of OHM fraction on 
\LFIR\ indicates that global properties are important in setting up the
right conditions for megamasers, but global properties are unlikely to 
determine the properties of the OHMs produced.  These notions are supported
by the poor trends of optical, FIR, and radio nuclear properties with 
OH emission line properties discussed below and in Paper V.  
The extra scatter introduced to the OH-FIR relation by folding in the
radio continuum data indicates that either OHMs are nearly saturated 
or that most of the radio continuum is emitted from regions which are not 
associated with masing (or both).

\subsection{The FIR-Radio Continuum Relationship}\label{sec:fir-radio}

\begin{figure}[!t]
%\centerline{\psfig{figure=Darling.fig10.epsi,clip=t,width=6in}}
\epsscale{1}
\plotone{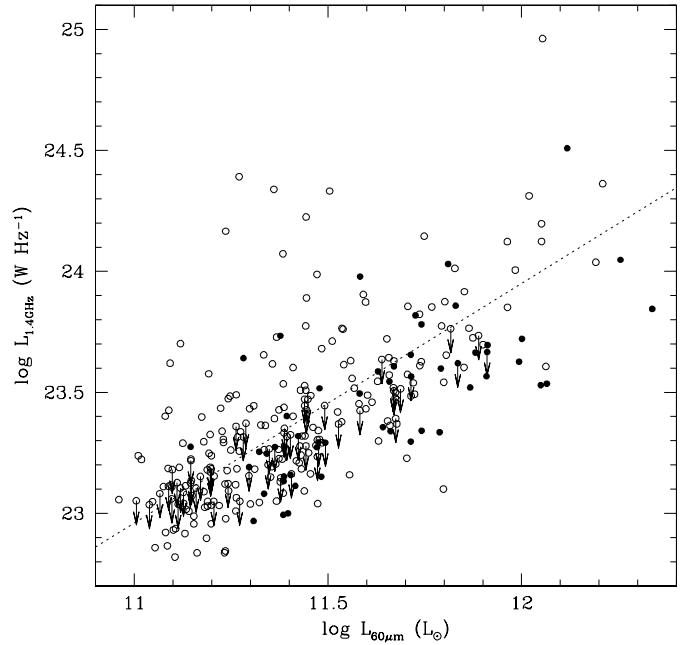}
\caption[The Radio-FIR Correlation of the Arecibo Survey]
{The radio-FIR correlation of the Arecibo survey.  The dotted line
is the radio-FIR correlation fit by Yun, Reddy, \& Condon (2001)
and is {\it not a fit} to the data.  Filled circles are OH megamasers
and open circles are OH nondetections.  Upper limits on $L_{1.4GHz}$
obtained from the NVSS are indicated by arrows.  
\label{radio-FIR}}
\end{figure}

A well-known relationship exists between the FIR and radio continuum
luminosities of star forming galaxies which spans roughly 5 orders of 
magnitude.  Yun, Reddy, \& Condon (2001) derive the form of the 
relationship from the 2 Jy {\it IRAS} sample (Strauss \etal 1992):
\begin{equation}
	\log L_{1.4GHz} = (0.99\pm0.01) \log L_{60\mu m} + (12.07\pm0.08)
  \label{eqn:radio-FIR}
\end{equation}
where $L_{1.4GHz}$ is in units of W Hz\minusone\ and $L_{60\mu m}$ is 
in solar luminosities.
Figure \ref{radio-FIR} presents the radio-60\micron\ relationship for
the Arecibo OHM survey, including OH nondetections.  The line on the 
plot is {\it not a fit} to the data;  it is Equation
\ref{eqn:radio-FIR}.  Figure \ref{radio-FIR} shows that there are some
radio ``monsters'' in the sample and that there appears to be a trend for 
OHMs to be radio under-luminous (or IR over-luminous) compared to the 
Arecibo survey LIRG sample population as a whole.
This may be a hint at the underlying properties of the interacting galaxies
which favor OHM production, but analysis is complicated by the 
upper limits on radio continuum luminosity.

\begin{figure}[!t]
%\centerline{\psfig{figure=Darling.fig11.epsi,clip=t,width=6in}}
\epsscale{1}
\plotone{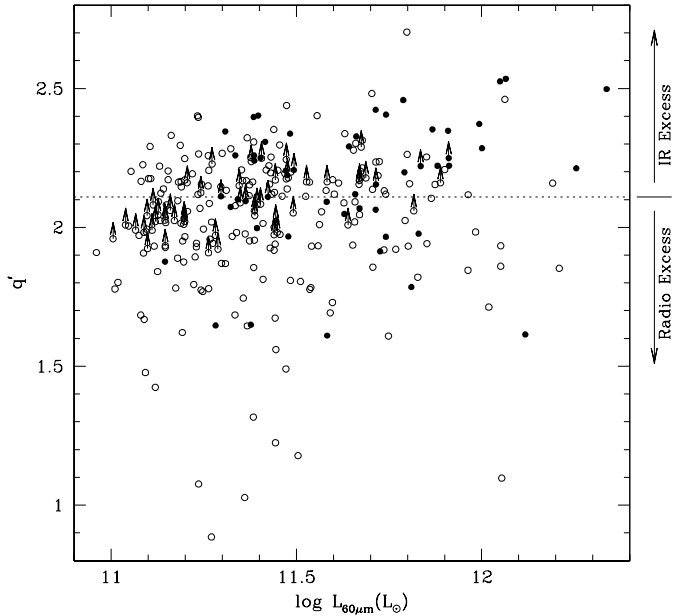}
\caption[Radio/IR Excess versus $L_{60\mu m}$]
{Radio/IR excess versus $L_{60\mu m}$.  The $q^\prime$ statistic is
a logarithmic measure of the 60\micron\ to radio flux density ratio
and is independent of distance.  Upper limits on radio flux density 
correspond to lower limits on $q^\prime$, as indicated by the arrows.
Filled circles are OH megamasers and open circles are OH nondetections.
The dotted horizontal line is $\overline{q^\prime} = 2.11$ derived from 
the fit of a large sample of star forming galaxies by Yun, Reddy, \&
Condon (2001).
\label{q_prime}}
\end{figure}

To more easily study departures of star forming galaxies from the FIR-radio
continuum relationship many groups examine the ``$q$'' parameter 
(Condon \etal 1991) which is a measure of the logarithmic FIR to radio flux 
density ratio:
\begin{equation}
	q \equiv \log\left(2.58 f_{60\mu m} + f_{100\mu m}\over 
	2.98\times10^{26}\ \mbox{Jy}\right) 
	- \log\left(f_{1.4GHz}\over\mbox{W m$^{-2}$ Hz$^{-1}$}\right). 
\end{equation}
We define a variation on the $q$ parameter 
called $q^\prime$ which does not depend on the {\it IRAS} 100 \micron\
flux:
\begin{equation}
	q^\prime \equiv \log\left(f_{60\mu m}\over 1.15\times10^{26}\ \mbox{Jy}\right) - \log\left(f_{1.4GHz}\over\mbox{W m$^{-2}$ Hz$^{-1}$}\right). 
	\label{eqn:q_prime}
\end{equation}
This version of the $q$ parameter is favored because many of the LIRGs in the
Arecibo OHM survey are not detected by {\it IRAS} at 100 \micron .  Also,
we can insert the radio-FIR relation of Equation \ref{eqn:radio-FIR} into
Equation \ref{eqn:q_prime} to 
obtain the average $q^\prime$ value versus $L_{60\mu m}$ and compare it to the
sample of OHMs and nondetections:
\begin{equation}
	\overline{q^\prime} = 0.01\log\left(L_{60\mu m}\over L_\odot\right) + 2.00.
\end{equation}
The Arecibo OHM survey data spans $L_{60\mu m} = 10^{11}$--$10^{12.4}L_\odot$,
over which $\overline{q^\prime}$ changes by 0.01, which is within the 
uncertainty
of the radio-FIR correlation.  We set $\overline{q^\prime} = 2.11$, 
the value at
$L_{60\mu m} = 10^{11.7} L_\odot$.  Figure \ref{q_prime} shows the 
distribution of $q^\prime$ values for the Arecibo survey sample, including
sources which were not detected by the NVSS and thus have lower bounds for
$q^\prime$.% (should we fill in with FIRST?).  

Cursory examination of Figure \ref{q_prime} indicates that OHMs may
as a population have an IR excess compared to the OH nondetections.
To include the upper limits on radio flux density, we perform a 
survival analysis.  Using survival analysis techniques 
available in the IRAF ASURV package  
Rev 1.2 (LaValley, Isobe, \& Feigelson 1992), which implements the 
methods presented in Feigelson \& Nelson (1985),
we obtain Kaplan-Meier estimates of the means and standard deviations
of the $q^\prime$ values for the OHMs of $2.19\pm0.03$ and for the
nondetections of $2.06\pm0.02$.  These appear to be statistically 
different populations among the LIRGs, with OHM hosts showing a IR
excess compared to the nondetections.  Rather than simply comparing
the means of the populations, one can use all of the information 
contained in the complete samples and compare the Kaplan-Meier maximum
likelihood estimator of the $q^\prime$ distribution of the two samples,
as shown in Figure \ref{q_prime:distribution}.  Also, we use ASURV
to compute non-parametric test statistics to determine the probability
that the two populations were drawn from the same underlying sample,
similar to the Kolmogorov-Smirnov test.  The variety of possible tests
stems from their vulnerability to different censoring distributions, and
a safe approach is to compute a variety of test statistics and compare
the results bearing in mind the nature of the censoring distribution in
the data.  The available test statistics are Gehan's generalized Wilcoxon
test (both permutation and hypergeometric variance), the logrank test,
the Peto \& Peto generalized Wilcoxon test, and the Pet \& Prentice
generalized Wilcoxon test (Feigelson \& Nelson 1985).
In this case, the censoring distribution is the same for
both populations, although it may not be completely random.  The 
two-population test statistics unanimously determine that the two 
populations have different $q^\prime$ distributions to high significance
(0.01--0.13$\%$ probability that the two populations are the same).  

This is a surprising result relating the star formation properties 
of LIRGs to the production of OHMs:  OH megamaser hosts have
a IR excess or a radio deficit compared to the OH nondetections
and the LIRG population overall.  This could either indicate that 
OHM hosts have buried AGN which make an extra contribution to the total
FIR budget, or that OHM hosts are undergoing a very recent/violent burst
of star formation in which the radio emission has not yet ``caught up.''
%(ref?? - I've seen something somewhere about this).  
In either case, 
this result is quite surprising because ostensibly OH emission is 
proportional to both FIR and radio luminosity (the pump and the stimulant,
respectively), but it appears that OHMs prefer hosts with an extra-strong
pump.  
The next logical question to ask is:  Are OHM hosts IR overluminous or
radio underluminous?

The same two-population survival analysis performed on the radio
luminosity of the OHMs and nondetections reveals a difference 
between the populations, but at a much less significant level.
The various test statistics converge on differing populations at
the 2.5$\%$ significance level, and the Kaplan-Meier estimator of
the mean value for $\log L_{1.4GHz}$ (W Hz\minusone) is $23.45\pm0.05$ 
for OHM hosts and $23.35\pm0.02$ for nondetections (the difference
between these is nonzero by only 2$\sigma$).  
Although there are no censored 60 \micron\ data points in the 
survey sample, we perform the identical survival analysis on the
60 \micron\ luminosity to compare to the results for $q^\prime$ and the
radio luminosity.  The relevant test statistics find a highly significant
($<0.01\%$) difference between the populations, and the Kaplan-Meier 
estimator of the mean value for $\log L_{60\mu m}$ ($L_\odot$) is
$11.64\pm0.04$ for the OHM hosts and $11.41\pm0.02$ for the nondetections.
These large differences in the FIR properties of OHM hosts from the
LIRG population overall were also illustrated in Figure \ref{survey_color} 
and discussed in \S \ref{sec:ohfrac}.  This preference for high FIR 
luminosity hosts appears to dominate the effect seen in the $q^\prime$ 
analysis, and indicates that the OHM hosts are not so much radio deficient
as they are IR abundant.  In other words, objects which are IR-overluminous
are more likely to produce OHMs independent of the radio emission properties.  
This is consistent with the nearly saturated masing state of OHMs
determined from the OH-FIR relation in \S \ref{sec:relation}.

\begin{figure}[!t]
%\centerline{\psfig{figure=Darling.fig12.epsi,clip=t,width=6in}}
\epsscale{1}
\plotone{Darling.fig11.epsi}
\caption[Estimated Distributions of the Radio/IR Excess of OHMs and OH 
Nondetections]
{Estimated distributions of the radio/IR excess of OHMs and OH nondetections.
The $q^\prime$ statistic is
a logarithmic measure of the 60\micron\ to radio flux density ratio
and is independent of distance.  $F(q^\prime)$ is the Kaplan-Meier survival
analysis estimate of the distribution of $q^\prime$, for which only lower
limits exist for some objects in the survey.  Shown are the distributions
and estimated uncertainties for OHMs (dotted line) and the nondetections
(solid line).  The vertical dashed line indicates 
$\overline{q^\prime} = 2.11$ derived from 
the fit of a large sample of star forming galaxies by Yun, Reddy, \&
Condon (2001).  The vertical arrows indicate the Kaplan-Meier estimates 
of the mean and standard deviation of the two populations.  
\label{q_prime:distribution}}
\end{figure}

\begin{figure}[!t]
%\centerline{\psfig{figure=Darling.fig13.epsi}}
\epsscale{1}
\plotone{Darling.fig12.epsi}
\caption[Radio/IR Excess versus \LOH]
{Radio/IR excess versus \LOH.
The $q^\prime$ statistic is
a logarithmic measure of the 60\micron\ to radio flux density ratio
and is independent of distance.  Upper limits on radio flux density 
correspond to lower limits on $q^\prime$, as indicated by the arrows.
The dotted horizontal line is $\overline{q^\prime} = 2.11$ derived from 
the fit of a large sample of star forming galaxies by Yun, Reddy, \&
Condon (2001).
\label{qp_LOH}}
\end{figure}

If LIRGs with an IR excess favor OHM production, does the excess also relate
to the properties of the OH emission?  Figure \ref{qp_LOH} shows the 
distribution of $q^\prime$ versus \LOH\ for the Arecibo OHM sample.  There
is no obvious correlation, again showing that the global properties of 
LIRGs can indicate the likelihood of OHM production, but the properties of the
OHMs produced tend to rely on smaller scale physics.

\begin{figure*}[!t]
%\centerline{\psfig{figure=Darling.fig14.epsi,clip=t,width=6in}}
\epsscale{1.5}
\plotone{Darling.fig13.epsi}
\caption[Relationships Between OH Line Properties]
{Relationships between OH line properties.
\label{OHline}}
\end{figure*}

\subsection{The Manifold of OH Megamasers and Their Hosts}

\subsubsection{Basic Relationships}

Few unambiguous relationships are evident among the properties of 
OHMs and their hosts.  In general, the scatter in any given 
parameter is large compared to measurement errors.  Figure 
\ref{OHline} plots various pairwise combinations of \LOH, $\Delta v$, 
$W_{1667}$, and $f_{OH}$.  The most luminous OHMs tend to have 
the highest peak flux densities but are not necessarily broad.  
Line width and peak show very little correlation.  One of the prominent
outliers in these plots is {\it IRAS} 02524+2046, an exceptionally 
narrow but exceptionally strong OHM.  

\begin{figure*}[ht!]
%\centerline{\psfig{figure=Darling.fig15.epsi,clip=t,width=6in}}
\epsscale{1.5}
\plotone{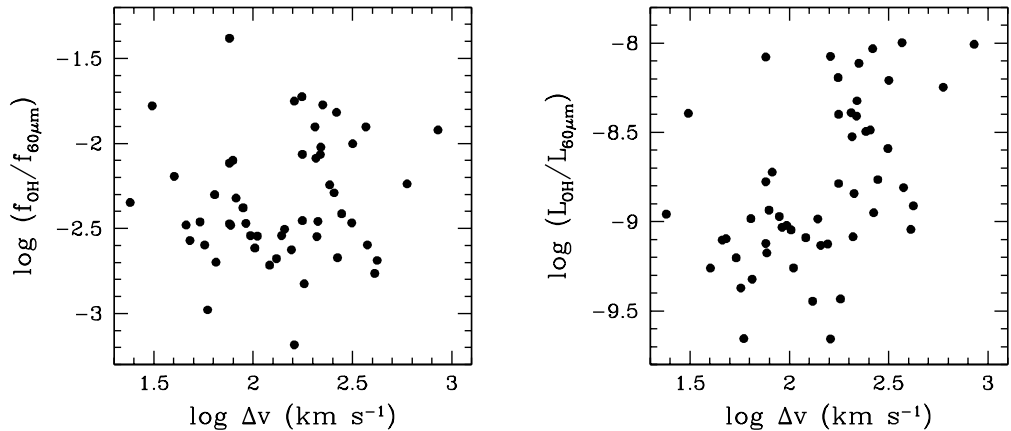}
\caption[OH Pumping Efficiency versus Line Width]
{OH pumping efficiency versus line width.  
Two measures of OH pumping efficiency are plotted:  the
ratio of the peak OH flux density to the 60 \micron\ flux density, and the
ratio of the OH line luminosity to the 60 \micron\ luminosity.  Both 
measures show poor correlation with OH line width, unlike the result 
obtained by Staveley-Smith \etal (1992) which shows an anticorrelation
(narrower OH lines have higher gain).
\label{OHeff}}
\end{figure*}

Two measures of OH pumping efficiency are plotted in Figure \ref{OHeff}:  the
ratio of the peak OH flux density to the 60 \micron\ flux density, and the
ratio of the OH line luminosity to the 60 \micron\ luminosity.  Both 
measures show poor correlation with OH line width, unlike the result 
obtained by Staveley-Smith \etal (1992) which shows an anticorrelation
(narrower OH lines have higher efficiency).
Note that if $L_{OH}/L_{60\mu m}$ is constant with respect to $\Delta v$, 
then $f_{OH}/f_{60\mu m} \propto \Delta v^{-1}$ if the integrated flux density
of the OH line obeys $F_{OH} \propto \Delta v\ f_{OH}$.  Alternatively, if
$f_{OH}/f_{60\mu m}$ is constant with respect to $\Delta v$, then 
$L_{OH}/L_{60\mu m} \propto \Delta v$.   Figure \ref{OHeff} shows that
the data favor the latter relationship, although we know from measuring
OH profiles that the integrated OH line flux density can show a large
variation for a given $\Delta v$ due to the menagerie of emission profiles
present in the OHM sample.  The varied nature of the OH line profiles
is likely to be the main source of the scatter in the relationship between
$L_{OH}/L_{60\mu m}$ and $\Delta v$.  

If the pumping efficiency measure 
$f_{OH}/f_{60\mu m}$ is constant with respect to $\Delta v$, what does 
this imply?  It may indicate that the line width is determined by 
the velocity structure in the gas produced by a merger while the 
frequency of maximum pumping efficiency is simply the velocity with the 
largest number of masers along the line of sight.  In this scenario, 
the observed line profiles are a combination of the complicated gas 
dynamics found in the nuclei of merging systems and projection effects.

\subsubsection{A Principal Component Analysis}

A principal component analysis is made of (nearly) the same 
properties analyzed by
Staveley-Smith \etal (1992):  $\log f_{100\mu m}/f_{60\mu m}$, 
$\log L_{60\mu m}$, 
$\log \Delta v$ (the rest frame velocity width), $\log L_{OH}$, and 
$\log L_{1.4GHz}$.  Minor differences reside in the measured radio 
luminosity (Staveley-Smith \etal define a radio luminosity spanning 0-10 GHz
based on flux densities measured at 5 GHz), our use of $L_{60\mu m}$ 
instead of \LFIR, and the correction of the OH line width to the 
OHM rest frame.  For 13 OHMs, Staveley-Smith \etal obtain a principal
plane which accounts for $93\%$ of the total variance in the sample
and which reduces the RMS residuals in all parameters except \LOH\ to 
the estimated observational errors.  Of the 52 OHMs in the Arecibo 
survey sample, 18 do not have measured {\it IRAS} 100 \micron\ fluxes
or were not detected in the NVSS, leaving 34 OHMs for this analysis.  
We used a principal component analysis code written by Murtagh, 
described by Murtagh \& Heck (1987), and provided by the multivariate index of 
StatLib\footnote{http://lib.stat.cmu.edu}, a service hosted by Carnegie 
Mellon University.
Table \ref{PC_1} presents the results of the principal component analysis
in the same manner as the Staveley-Smith \etal analysis for ease of 
comparison, including the correlation matrix, all of the eigenvectors and
eigenvalues, all of the RMS residuals for progressively higher order fits, 
and the estimated observational errors in each parameter.  

The correlation matrix indicates positive relationships between 
the OH, FIR, and radio luminosities as well as between the OH line
width and the OH and radio luminosities.  Recall that the luminosities
are correlated in part due to Malmquist bias.  FIR color exhibits only 
weak anti-correlations with the other parameters.
Comparison of the RMS residuals to the observational errors indicates 
that the data set is not well described by a principal axis or principal
plane.  Only a 3-plane begins to bring the RMS residuals down to the
level of the observation errors with the exception of \LOH.  A similar 
extra variance in \LOH\ is seen by Staveley-Smith \etal (1992) which 
they interpret to indicate that OH masing relies on hidden variables
such as beaming.
The first few eigenvectors are not terribly
illuminating either.  The first principal component,
which accounts for $56\%$ of the variance, gets a uniform contribution from
all variables except the color term.  The second component draws mostly
from the FIR color and the OH line width and accounts for $20\%$ of the
variance.  The third component gets a nearly uniform contribution from
all parameters but with the width and \LOH\ in the opposite sense from
the rest of the terms.  

Analysis of other combinations of parameters related to OHMs and their hosts
does not reveal any new information or trends.  In general, the observation
errors are small compared to the scatter in any given parameter.  The 
residuals in \LOH\ are always among the most difficult to reduce through 
projections onto principal axes, indicating that there are hidden variables
influencing the properties of OHMs.  This is hardly surprising, given the
size scales of most of the properties we measure compared to the size scales
of masing.  Factors likely to affect \LOH\ include beaming, saturation 
state(s), and collisional pumping.

\section{Conclusions}

All of these analyses are folded into
a new understanding of the OH megamaser phenomenon which unfortunately tends
to reject the conventional wisdom about OHM environments and production
mechanisms without offering a clean new picture.  Most of the difficulty
with constructing a model for OHM production lies in the large scatter
intrinsic to the masing amplification process which relies on small 
scale conditions found in largely unresolved sources.  
OHMs show significant intrinsic variation in properties, producing 
intrinsic scatter roughly equal to the span of the sample data.  
Higher resolution
imaging and spectroscopy in the radio, optical, and infrared bands
offers some promise for placing the OHMs in the context of a merging 
systems and relating OHM activity to the stage and activity level 
of major galaxy mergers.  Paper V touches on a few of these issues and
points towards fairly short lifetimes for OHMs: certainly less than 
$10^8$ years and more likely less than $10^7$ years.  

We have demonstrated the ability to identify OHM candidates with much
success by selecting systems with high \LFIR.  
FIR-selected candidates at $z>0.1$ conspire with the increasing OHM 
fraction versus \LFIR\ to produce a high detection rate (1 in 5.5), 
doubling the sample of OHMs overall, and increasing the sample at
$z>0.1$ sevenfold.  The survey detections are incorporated into a
reliable OH luminosity function and projected to high redshift for 
various merger evolution scenarios in Paper IV\@.  It is likely that
OH gigamasers (if they exist) may be detected out to $z=5$ with future 
instrumentation, and can serve as luminous radio
tracers of merging systems, dust-obscured star formation, and 
the formation of binary supermassive black holes.  

Incorporating all reliable OHM detections, we reexamine the OH-FIR 
relationship and find:
$L_{OH} \propto L_{FIR}^{1.2\pm0.1}$.  This may indicate a mixture
of saturation states across the sample or within individual merging
systems.  It is likely that cases of mostly saturated and mostly 
unsaturated masing are in the Arecibo OHM sample, but spatially 
resolved spectral line maps would be required to confirm our
suspicions about individual systems.  

Significant trends between properties of OHMs and their hosts or
between OHM line properties are few.  There is generally a mismatch
of size scales between masing, which amplifies small-scale conditions, 
and properties of mostly unresolved merging systems, which represent
integrated global quantities.  The trend of increasing OHM fraction 
with increasing \LFIR\ and warmer FIR colors indicates that global
properties of merging systems can indicate the likelihood that a 
given merging system will host the environments which produce OHMs, 
but these global properties cannot predict the nature of the OHM 
which is produced due to the amplification of small-scale conditions.

The most luminous OHMs tend to be not only strong (with a high 
observed flux density) but broad, spanning more than 1000 km s\minusone\ 
at $10\%$ of peak flux density.  The two OH gigamasers detected by
this survey ({\it IRAS} F12032+1707 and F14070+0525) 
show multiple strong OH line components and are produced
in two of the most luminous ULIRGs in the survey sample.

\acknowledgements
 
The authors are very grateful to Will Saunders for access to the PSCz catalog
and to the excellent staff of NAIC for observing assistance and support.  
We thank the anonymous referees for insightful comments and constructive 
criticism.  
This research was supported by Space Science Institute archival grant 
8373 and NSF grant AST 00-98526
and made use of the NASA/IPAC Extragalactic Database (NED) 
which is operated by the Jet Propulsion Laboratory, California
Institute of Technology, under contract with the National Aeronautics 
and Space Administration.  
We acknowledge the use of NASA's SkyView facility 
(http://skyview.gsfc.nasa.gov) located at NASA Goddard Space Flight Center.

\begin{deluxetable}{cccccrrrrcc}
\tabletypesize{\scriptsize}
\tablecaption{OH Non-Detections:  Optical Redshifts and FIR Properties\label{nondetectFIR}}
\tablewidth{0pt}
\tablehead{
\colhead{{\it IRAS} Name} &  \colhead{$\alpha$} & \colhead{$\delta$} & \colhead{$z_\odot$}  
& \colhead{Ref}& \colhead{$v_\odot$} & \colhead{$v_{CMB}$} 
& \colhead{$D_L$} & \colhead{$f_{60}$} & \colhead{$f_{100}$} & 
\colhead{$\log L_{FIR}$} \\
\colhead{FSC} & \colhead{B1950} & \colhead{B1950} &  & \colhead{} & \colhead{km/s} &
\colhead{km/s} &  \colhead{$h^{-1}_{75}$ Mpc} &  \colhead{Jy} &  
\colhead{Jy} & \colhead{$h^{-2}_{75} L_\odot$} \\
\colhead{(1)}& \colhead{(2)}& \colhead{(3)}& \colhead{(4)}& \colhead{(5)}& 
\colhead{(6)}& \colhead{(7)}& \colhead{(8)}& \colhead{(9)}& \colhead{(10)}& 
\colhead{(11)}
 }
\startdata
00020+3636 & 00 02 06.4 & +36 37 01 & 0.1179 & 1 & 35344(158)
& 35031(159) & 494(2) & 0.857(\phn77) & 0.97(23) & 11.49 \\
00085+3107 & 00 08 32.0 & +31 07 14 & 0.1025 & 1 & 30730(102)
& 30404(104) & 426(2) & 0.693(\phn55) & 0.90(18) & 11.28 \\
00109+0228 & 00 10 57.6 & +02 28 31 & 0.1590 & 1 & 47672(117)
& 47318(117) & 681(2) & 0.648(104) & 1.32(17) & 11.74 \\
00128+2817 & 00 12 47.1 & +28 17 18 & 0.1405 & 1 & 42108(149)
& 41777(150) & 596(2) & 0.753(\phn60) & 1.12(16) & 11.63 \\
00268+2413 & 00 26 53.2 & +24 13 38 & 0.1094 & 1 & 32790(121)
& 32457(122) & 456(2) & 1.221(\phn86) & 1.42(17) & 11.58 \\
00310+1437 & 00 31 01.9 & +14 37 09 & 0.1154 & 1 & 34600(\phn29)
& 34257(\phn32) & 483(0) & 0.992(\phn79) & 1.57(24) & 11.58 \\
00331+2656 & 00 33 11.3 & +26 56 15 & 0.1923 & 1 & 57644(135)
& 57319(136) & 837(2) & 0.703(\phn63) & 1.48(19) & 11.96 \\
00486+0004 & 00 48 41.6 & +00 04 45 & 0.1127 & 1 & 33781(\phn55)
& 33449(\phn57) & 471(1) & 0.610(\phn61) & 0.97(22) & 11.35 \\
\enddata
\tablerefs{Redshifts were obtained from:  
(1) Saunders \etal 2000; 
(2) Lawrence \etal 1999; 
(3) Klaas \& Els\"{a}sser 1991; 
(4) Huchra \etal 1983; 
(5)  Fisher \etal 1995;  
(6) Kim \& Sanders 1998; 
(7) Clowes \etal 1995;
(8) Leech \etal 1994;
(9) Kim \etal 1995; 
(10) Downes, Solomon, \& Radford 1993; 
(11) Strauss \& Huchra 1988;  
(12) Beers \etal 1995; 
(13) Dey, Strauss, \& Huchra 1990.}
\end{deluxetable}

\begin{deluxetable}{cccccrrrrc}
\tabletypesize{\scriptsize}
\tablecaption{OH Non-Detections:  OH Limits and 1.4 GHz Properties\label{nondetectOH}}
\tablewidth{0pt}
\tablehead{
\colhead{{\it IRAS} Name} &  \colhead{$z_\odot$}  
& \colhead{$\log L_{FIR}$} & \colhead{$\log L^{pred}_{OH}$} 
& \colhead{$\log L^{max}_{OH}$} & \colhead{$t_{on}$} & \colhead{RMS} 
& \colhead{$f_{1.4GHz}$\tablenotemark{a}} & \colhead{Class\tablenotemark{b}} 
& \colhead{Note} \\
\colhead{FSC} &  & \colhead{$h^{-2}_{75} L_\odot$} &
\colhead{$h^{-2}_{75} L_\odot$} & \colhead{$h^{-2}_{75} L_\odot$} & \colhead{min} 
& \colhead{mJy} & \colhead{mJy} & \\
\colhead{(1)}& \colhead{(2)}& \colhead{(3)}& \colhead{(4)}& \colhead{(5)}& 
\colhead{(6)}& \colhead{(7)}& \colhead{(8)}& \colhead{(9)}& \colhead{(10)}
 }
\startdata
00020+3636 & 0.1179 & 11.49 & 1.83 & 1.80 & 12 & 0.74 & 15.4(1.4) & \\
00085+3107 & 0.1025 & 11.28 & 1.55 & 1.52 & 12 & 0.51 & \<\phn5.0\phn& \\
00109+0228 & 0.1590 & 11.74 & 2.18 & 2.02 & 12 & 0.67 & \<\phn5.0\phn& \\
00128+2817 & 0.1405 & 11.63 & 2.03 & 1.84 & 16 & 0.56 & 4.9(0.5) & \\
00268+2413 & 0.1094 & 11.58 & 1.95 & 1.45 & 40 & 0.38 & 6.6(0.5) & \\
00310+1437 & 0.1154 & 11.58 & 1.96 & 1.65 & 12 & 0.54 & \<\phn5.0\phn& & 1 \\
00331+2656 & 0.1923 & 11.96 & 2.49 & 2.20 & 12 & 0.68 & 8.5(0.6) & L(1)\\
00486+0004 & 0.1127 & 11.35 & 1.64 & 1.73 & 12 & 0.69 & \<\phn5.0\phn& &2 \\
\enddata
\tablenotetext{a}{1.4 GHz continuum fluxes are courtesy of the NRAO VLA Sky Survey 
(Condon \etal 1998).}
\tablenotetext{b}{Spectral classifications use the codes:  
``S2'' = Seyfert type 2;  ``S1'' = Seyfert type 1;
``H'' = \HII region (starburst);  
and ``L''~=~low-ionization emission region (LINER).  }
\tablerefs{Spectral classifications were obtained from: 
(1) Darling \& Giovanelli 2002b (Paper V); 
(2) Wu \etal 1998; 
(3) Veilleux, Kim, \& Sanders 1999; 
(4) Lawrence \etal 1999; 
(5) Kim, Veilleux, \& Sanders 1998; 
(6) Wilman \etal 1998; 
(7) Veilleux \etal 1995;
(8) Clowes \etal 1995; 
(9) Leech \etal 1994; 
(10) Moran, Halpern, \& Helfand 1996; 
(11) White \etal 2000.}
\tablecomments{
(1) Source needs more integration time, due
to a suggestive feature in the bandpass; 
(2) Source needs more integration time, based on
$L^{pred}_{OH} < L^{max}_{OH}$; 
(3) Lu \& Freudling (1995) list an HI redshift and spectrum with $z=0.04714$ which is
not unambiguous, but casts doubt on the PSCz redshift of this source;
(4) Galactic \HI in bandpass; 
(5) Possible OH absorption feature, but $\lesssim1$ mJy deep;
(6) Radio quiet IR QSO (Beichman \etal 1986);
(7) Observations were performed during daylight, which increases the 
RMS noise significantly; 
(8) Possible 2--3 mJy OH emission feature 220 km s\minusone\ 
redward of the optical redshift.}
\end{deluxetable}

\begin{deluxetable}{lccccrrrrccc}
\tabletypesize{\scriptsize}
\tablecaption{Ambiguous Candidates:  Optical Redshifts and FIR Properties\label{ambig:opt}}
\tablewidth{0pt}
\tablehead{
\colhead{{\it IRAS} Name} &  
\colhead{$\alpha$} & 
\colhead{$\delta$} & 
\colhead{$z_\odot$} & 
\colhead{Ref} &
\colhead{$v_\odot$} & 
\colhead{$v_{CMB}$} & 
\colhead{$D_L$} &
\colhead{$f_{60\mu m}$} &
\colhead{$f_{100\mu m}$} &
\colhead{$\log L_{FIR}$} &
\colhead{Note}\\
\colhead{FSC} & \colhead{B1950} & \colhead{B1950} &\colhead{} &
\colhead{z} & \colhead{km/s} & \colhead{km/s} & \colhead{$h^{-1}_{75}$Mpc} & 
\colhead{Jy} & \colhead{Jy} & \colhead{$h^{-2}_{75} L_\odot$} \\
\colhead{(1)}&\colhead{(2)}&\colhead{(3)}&\colhead{(4)}&\colhead{(5)}&
\colhead{(6)}&\colhead{(7)}&\colhead{(8)}&\colhead{(9)}&\colhead{(10)}&
\colhead{(11)} & \colhead{(12)}
}
\startdata
00242+3344 & 00 24 14.9 & +33 44 51 & 0.1737 & 1 & 52067(135)
& 51754(137) & 750(2) & 0.601(\phn84) & 1.33(23) & 11.81 & 1\\
00302+3625 & 00 30 17.7 & +36 25 26 & 0.3023 & 1 & 90628(159)
& 90325(161) & 1386(3) & 0.682(\phn61) & 0.77(17) & 12.28 & 2\\
01348+3254 & 01 34 48.8 & +32 54 07 & 0.3670 & 2 & 110024(\phn\phn\phn)
& 109752(\phn27) & 1731(0) & 0.740(\phn52) & \<\phn1.15 & 12.36--12.56 & 3,4\\
01506+2554 & 01 50 39.2 & +25 54 54 & 0.3264 & 1 & 97855(381)
& 97584(382) & 1513(7) & 0.578(\phn52) & 1.14(21) & 12.38 & 2\\
02054+0835 & 02 05 27.5 & +08 35 54 & 0.3450 & 3 & 103428(250)
& 103161(251) & 1612(5) & 0.561(\phn67) & \<\phn1.99 & 12.17--12.55 & 2\\
06587+3043 & 06 58 47.8 & +30 43 42 & 0.1727 & 1 & 51788(121)
& 51910(127) & 752(2) & 0.673(\phn74) & \<\phn1.98 & 11.59--11.92 & 1\\
07488+0501 & 07 48 52.4 & +05 01 17 & 0.1732 & 1 & 51935(129)
& 52165(132) & 756(2) & 0.661(\phn53) & 1.18(11) & 11.82 & 1\\
08519+2017 & 08 51 58.9 & +20 17 57 & 0.3060 & 1 & 91738(\phn\phn\phn)
& 92006(\phn27) & 1415(0) & 0.891(\phn62) & 1.16(23) & 12.44 & 2,3\\
\enddata
\tablenotetext{a}{This marginal OH detection (3.4$\sigma$; Dickey et al 1990) 
was undetectable at Arecibo due to standing waves produced by its 
strong continuum.}
\tablerefs{Redshifts were obtained from:  
(1) Saunders \etal 2000; 
(2) Spinrad \etal 1985;
(3) Lawrence \etal 1999; 
(4) Strauss \etal 1992; 
(5) Leech \etal 1994; 
(6) Nakanishi \etal 1997. }
\tablecomments{
(1) 1667 MHz OH line undetectable due to Galactic \HI; 
(2) OH undetectable due to RFI; 
(3) Continuum creates standing waves, making OH undetectable;
(4) 3C048; 
(5) 3C273;
(6) Continuum source (444 mJy; Condon \etal 1998) 
82\arcsec from this target creates standing waves, making OH undetectable;
(7) 4C12.50;
(8) 3C459.}
\end{deluxetable}

\begin{deluxetable}{ccccrcc} 
\tabletypesize{\scriptsize}
\tablecaption{Ambiguous Candidates:  1.4 GHz Properties and Notes
	\label{ambig:radio}}
\tablewidth{0pt}
\tablehead{
\colhead{{\it IRAS} Name} &  \colhead{$z_\odot$}  
& \colhead{$\log L_{FIR}$} & \colhead{$\log L^{pred}_{OH}$} 
& \colhead{$f_{1.4GHz}$\tablenotemark{a}} & \colhead{Class\tablenotemark{b}} 
& \colhead{Note} \\
\colhead{FSC} &  & \colhead{$h^{-2}_{75} L_\odot$} &
\colhead{$h^{-2}_{75} L_\odot$} 
& \colhead{mJy} & \\
\colhead{(1)}& \colhead{(2)}& \colhead{(3)}& \colhead{(4)}& \colhead{(5)}& 
\colhead{(6)}& \colhead{(7)}
}
\startdata
00242+3344 & 0.1737 & 11.81 & 2.27 & 20.1(0.8) & & 1\\
00302+3625 & 0.3023 & 12.28 & 2.93 & \<\phn5.0\phn & & 2\\
01348+3254 & 0.3670 & 12.36--12.56 & 3.03--3.31 &16018(481)& Q & 3,4\\
01506+2554 & 0.3264 & 12.38 & 3.06 & 40.3(1.6) & & 2\\
02054+0835 & 0.3450 & 12.17--12.55 & 2.78--3.30 & 10.5(1.2) & S1(1)& 2\\
06587+3043 & 0.1727 & 11.59--11.92 & 1.97--2.43 & 6.8(0.5) & & 1\\
07488+0501 & 0.1732 & 11.82 & 2.29 & 3.4(0.5) & & 1\\
08519+2017 & 0.3060 & 12.44 & 3.14 & 1512.3(45.4) & Q & 2,3\\
\enddata
\tablenotetext{a}{1.4 GHz continuum fluxes are courtesy of the NRAO VLA Sky Survey 
(Condon \etal 1998).}
\tablenotetext{b}{Spectral classifications use the codes:  
``S2'' = Seyfert type 2;  ``S1'' = Seyfert type 1; ``Q''~=~Quasar; and
``L''~=~low-ionization emission region (LINER).  }
\tablenotetext{c}{This marginal OH detection (3.4$\sigma$; Dickey et al 1990) 
was undetectable at Arecibo due to standing waves produced by its 
strong continuum.}
\tablerefs{Spectral classifications were obtained from: 
(1) Lawrence et al 1999; 
(2) Leech et al 1994;
(3) Kim, Veilleux, \& Sanders 1998; 
(4) Darling \& Giovanelli 2002b (Paper V);
(5) Veilleux, Kim, \& Sanders 1999.}
\tablecomments{
(1) 1667 MHz OH line undetectable due to Galactic \HI; 
(2) OH undetectable due to RFI; 
(3) Continuum creates standing waves, making OH undetectable;
(4) 3C048; 
(5) 3C273;
(6) Continuum source (444 mJy; Condon \etal 1998) 
82\arcsec from this target creates standing waves, making OH undetectable;
(7) 4C12.50;
(8) 3C459.}
\end{deluxetable}

\begin{deluxetable}{rccccrrrrcc}
\tabletypesize{\scriptsize}
\tablecaption{OH Detections:  Optical Redshifts and FIR Properties 
		\label{detectFIR}}
\tablewidth{0pt}
\tablehead{
\colhead{{\it IRAS} Name} &  \colhead{$\alpha$} & \colhead{$\delta$} & 
\colhead{$z_\odot$} & 
\colhead{Ref} &
\colhead{$v_\odot$} & 
\colhead{$v_{CMB}$} & 
\colhead{$D_L$} &
\colhead{$f_{60\mu m}$} &
\colhead{$f_{100\mu m}$} &
\colhead{$\log L_{FIR}$} 
 \\
\colhead{FSC} & \colhead{B1950} & \colhead{B1950} &\colhead{} &
\colhead{z} & \colhead{km/s} & \colhead{km/s} & \colhead{$h^{-1}_{75}$Mpc} & 
\colhead{Jy} & \colhead{Jy} & \colhead{$h^{-2}_{75} L_\odot$} \\
\colhead{(1)}&\colhead{(2)}&\colhead{(3)}&\colhead{(4)}&\colhead{(5)}&
\colhead{(6)}&\colhead{(7)}&\colhead{(8)}&\colhead{(9)}&\colhead{(10)}&
\colhead{(11)}
}
\startdata
01562+2528 & 01 56 12.0 & +25 27 59 & 0.1658 & 1 & 49707(505)
& 49441(506) & 714(8) & 0.809(\phn57) & 1.62(24) & 11.87 \\
02524+2046 & 02 52 26.8 & +20 46 54 & 0.1815 & 1 & 54421(125)
& 54213(129) & 788(2) & 0.958(\phn77) & \<\phn4.79 & 11.78--12.25 \\
03521+0028 & 03 52 08.5 & +00 28 21 & 0.1522 & 2 & 45622(138)
& 45501(142) & 653(2) & 2.638(237) & 3.83(34) & 12.25 \\
03566+1647 & 03 56 37.8 & +16 47 57 & 0.1335 & 1 & 40033(\phn54)
& 39911(\phn65) & 568(1) & 0.730(\phn66) & \<\phn2.37 & 11.38--11.73 \\
10035+2740 & 10 03 36.7 & +27 40 19 & 0.1662 & 1 & 49826(300)
& 50116(301) & 724(5) & 1.144(126) & 1.63(161) & 11.98 \\
10378+1108\tablenotemark{a} & 10 37 49.1 & +11 09 08 & 0.1362 & 2 & 40843(\phn61)
& 41190(\phn62) & 587(1) & 2.281(137) & 1.82(18) & 12.02 \\
11180+1623\tablenotemark{b}& 11 18 06.7 & +16 23 16 & 0.1660 & 3 & 49766(\phn70)
& 50104(\phn71) & 724(1) & 1.189(\phn95) & 1.60(18) & 11.99 \\
12005+0009 & 12 00 30.2 & +00 09 24 & 0.1226 & 1 & 36759(177)
& 37116(177) & 526(3) & 0.736(\phn88) & 0.98(15) & 11.50 \\
\enddata
\tablenotetext{a}{A known OHM included in the survey sample.}
\tablenotetext{b}{{\it IRAS} 11180+1623 is not in the PSCz catalog (excluded 
by the PSCz mask; Saunders \etal 2000).  It is in the 1 Jy survey (Kim \& Sanders 1998).}
\tablenotetext{c}{{\it IRAS} 12107+3157 is an OH absorber.}
\tablerefs{Redshifts were obtained from:  
(1) Saunders \etal 2000; 
(2) Strauss \etal 1992;
(3) Kim \& Sanders 1998; 
(4) Fisher \etal 1995; 
(5) Lawrence \etal 1999.}
\end{deluxetable}

\begin{deluxetable}{cccrcccrcccr}
\tabletypesize{\scriptsize}
\tablecaption{OH Detections:  OH Line and 1.4 GHz Continuum Properties \label{detectOH}}
\tablewidth{0pt}
\tablehead{
\colhead{{\it IRAS} Name} &  
\colhead{$v_{1667,\odot}$} & 
\colhead{$t_{on}$} & 
\colhead{$f_{1667}$} & 
\colhead{$W_{1667}$} &
\colhead{$\Delta \nu_{1667}$\tablenotemark{a}} & 
\colhead{$\Delta v_{1667}$\tablenotemark{b}} & 
\colhead{$R_H$} & 
\colhead{$\log L_{FIR}$} & 
\colhead{$\log L^{pred}_{OH}$} & 
\colhead{$\log L_{OH}$} &
\colhead{$f_{1.4GHz}$\tablenotemark{c}}  \\
\colhead{FSC} &\colhead{km/s} & \colhead{min} & \colhead{mJy} & \colhead{MHz}
& \colhead{MHz} & \colhead{km/s} &\colhead{} & \colhead{$h^{-2}_{75} L_\odot$} 
& \colhead{$h^{-2}_{75} L_\odot$} & \colhead{$h^{-2}_{75} L_\odot$} & \colhead{mJy} \\
\colhead{(1)}&\colhead{(2)}&\colhead{(3)}&\colhead{(4)}&\colhead{(5)}&
\colhead{(6)}&\colhead{(7)}&\colhead{(8)}&\colhead{(9)}&\colhead{(10)}&
\colhead{(11)}&\colhead{(12)}
}
\startdata
01562+2528 & 49814(15) & 32 & 6.95 & 1.29 & 1.04 & 218 & 5.9 & 11.87 & 2.37 & 3.22 & 6.3(0.5) \\
02524+2046 & 54162(15) & 24 & 39.82 & 0.50 & 0.36 &\phn76 & 3.2 & 11.78--12.25 & 2.24--2.89 & 3.71 & 2.9(0.6)\\
03521+0028 & 45512(14) & 72 & 2.77 & 0.61 & 0.29 &\phn59 & 5.8 & 12.25 & 2.89 & 2.41 & 6.7(0.6) \\
03566+1647 & 39865(14) & 84 & 1.96 & 0.98 & 0.23 &\phn48 & $\gtrsim\phn9.6$ & 11.38--11.73 & 1.69--2.17 & 2.29 & 3.5(0.5) \\
10035+2740 & 50065(14) & 76 &  2.29 & 0.74 & 0.31 &\phn65&$\sim15.4$& 11.98 & 2.51 & 2.47 & 6.3(0.5)\\
10378+1108\tablenotemark{d} & 40811(14) & 12 & 19.70 & 1.50 & 0.87 & 177 &\nodata& 12.02 & 2.57 & 3.51 & 8.9(0.6)\\
11180+1623\tablenotemark{e}& 49783(14) & 48 &  1.82 & 0.42 & 0.61 & 127 &$\geq\phn5.1$& 11.99 & 2.52 & 2.31 & 4.2(0.5) \\
12005+0009 & 36472(14) & 40 &  3.51 & 0.71 & 0.41 &\phn82&$\sim\phn2.0$& 11.50 & 1.85 & 2.60 & 5.4(0.6) \\
\enddata
\tablenotetext{a}{$\Delta \nu_{1667}$ is the {\it observed} FWHM.}
\tablenotetext{b}{$\Delta v_{1667}$ is the {\it rest frame} FWHM.  The rest
frame and observed widths are related by 
$\Delta v_{rest} = c(1+z)(\Delta\nu_{obs}/\nu_\circ)$.}
\tablenotetext{c}{1.4 GHz continuum fluxes are courtesy of the NRAO VLA Sky Survey 
(\cite{con98}).}
\tablenotetext{d}{Redetection of a known OHM included in the survey sample.}
\tablenotetext{e}{{\it IRAS} 11180+1623 is not in the PSCz catalog (excluded 
by the PSCz mask; Saunders \etal 2000).  
It is in the 1 Jy survey (Kim \& Sanders 1998).}
\tablenotetext{f}{{\it IRAS} 12107+3157 is an OH absorber.}
\end{deluxetable}

\begin{deluxetable}{lccccrrcrrc} 
\tabletypesize{\scriptsize}
\tablecaption{Known OH Megamasers:  Optical Redshifts and FIR Properties\label{OHM:opt}}
\tablewidth{0pt}
\tablehead{
\colhead{Name} &  \colhead{$\alpha$} & \colhead{$\delta$} & \colhead{$z_\odot$}  
& \colhead{Ref}& \colhead{$v_\odot$} & \colhead{$v_{CMB}$} 
& \colhead{$D_L$} & \colhead{$f_{60}$} & \colhead{$f_{100}$} & 
\colhead{$\log L_{FIR}$} \\
\colhead{FSC} & \colhead{J2000} & \colhead{J2000} &  & \colhead{} & \colhead{km/s} &
\colhead{km/s} &  \colhead{$h^{-1}_{75}$ Mpc} &  \colhead{Jy} &  
\colhead{Jy} & \colhead{$h^{-2}_{75} L_\odot$} \\
\colhead{(1)}& \colhead{(2)}& \colhead{(3)}& \colhead{(4)}& \colhead{(5)}& 
\colhead{(6)}& \colhead{(7)}& \colhead{(8)}& \colhead{(9)}& \colhead{(10)}& 
\colhead{(11)}
}
\startdata
%IRAS Name   &     RA	 &    Dec    & z_hel  &Ref& v_hel      & v_cmb     &  D     & f_60       & f_100      & L_FIR
%	    &   B1950	 &   B1950   &	      &   & km/s       & km/s      & Mpc    & (Jy)       & (Jy)       & (log)
00057+4021 & 00 05 44.9 & +40 21 15 & 0.0447 & 1 & 13389(\phn\phn9) & 13090(\phn24) & \phn178(0) & 4.47(0.36) & 4.30(0.30) & 11.30\\
00335$-$2732 & 00 33 32.0 & $-$27 32 05 & 0.0693 &2 & 20771(\phn36) & 20489(\phn42) & \phn283(1) & 4.29(0.47) & 3.21(0.22) & 11.65\\
00509+1225 & 00 50 57.8 & +12 25 20 & 0.0611 &3 & 18330(\phn20) & 17997(\phn26) & \phn247(0) & 2.24(0.18) & 2.63(0.26) & 11.31\\
01364$-$1042 & 01 36 24.3 & $-$10 42 24 & 0.0484 &4 & 14520(\phn81) & 14240(\phn84) & \phn194(1) & 6.16(0.62) & 6.70(0.54) & 11.53\\
01417+1651 & 01 41 47.4 & +16 51 02 & 0.0274 &2&  8245(\phn\phn5) &  7956(\phn25) & \phn107(0) &12.55(1.00) &13.31(0.93) & 11.32\\
02483+4302 & 02 48 19.8 & +43 02 53 & 0.0514 & 1 & 15421(\phn\phn9) & 15236(\phn35) & \phn208(0) & 4.02(0.28) & 6.92(0.41) & 11.47\\
03056+2034 & 03 05 38.6 & +20 34 54 & 0.0272 &6 &  8164(\phn36) &  7972(\phn49) & \phn108(1) & 4.41(0.26) & 5.89(0.65) & 10.90\\
03260$-$1422 & 03 26 04.6 & $-$14 22 25 & 0.0425 &2 & 12736(\phn55) & 12593(\phn64) & \phn171(1) & 2.91(0.29) & 4.84(0.34) & 11.16\\
\enddata
\tablerefs{Redshifts were obtained from:  
(1) Downes, Solomon, \& Radford 1993; 
(2) Strauss \etal 1992; 
(3) Solomon \etal 1997;
(4) Kim \etal 1995; 
(6) Huchra, Vogeley, Geller 1999;
(7) Sanders \etal 1995;
(8) Theureau \etal 1998;
(9) Kim \& Sanders 1998; 
(10) Smith \etal 1987;
(11) Nordgren \etal 1997;
(12) de Vaucouleurs \etal 1991;
(13)  Fisher \etal 1995;  
(14) Richter \& Huchtmeir 1987;
(15) Carilli, Wrobel, \& Ulvestad 1998;
(16) Haynes \etal 1997;
(17) Kartgert \etal 1998;
(18) Szomoru, van Gorkom, \& Gregg 1996;
(19) Strauss \& Huchra 1988;  
(20) Grogin \& Geller 2000;
(21) Reshetnikov \& Combes 1994;
(22) Saunders \etal 2000;
(23) Giovanelli \& Haynes 1993.
}
\end{deluxetable}

\begin{deluxetable}{lccccccrcc} 
\tabletypesize{\scriptsize}
\tablecaption{Known OH Megamasers:  OH and 1.4 GHz Properties\label{OHM:rad}}
\tablewidth{0pt}
\tablehead{
\colhead{{\it IRAS} Name} &  \colhead{$z_\odot$}  
& \colhead{$\log L_{FIR}$} & \colhead{$\log L^{pred}_{OH}$} 
& \colhead{$\log L_{OH}$} & \colhead{$f_{1667}$}  & \colhead{Ref}
& \colhead{$f_{1.4GHz}$\tablenotemark{a}} & \colhead{Class\tablenotemark{b}} 
& \colhead{Note} \\
\colhead{FSC} &  & \colhead{$h^{-2}_{75} L_\odot$} &
\colhead{$h^{-2}_{75} L_\odot$} & \colhead{$h^{-2}_{75} L_\odot$} & \colhead{mJy} 
 & \colhead{} & \colhead{mJy} & \colhead{}  \\
\colhead{(1)}& \colhead{(2)}& \colhead{(3)}& \colhead{(4)}& \colhead{(5)}& 
\colhead{(6)}& \colhead{(7)}& \colhead{(8)}& \colhead{(9)} & \colhead{(10)}
}
\startdata
%IRAS Name   &  z_hel & L_FIR & pred & LOH  & f_OH    & NVSS     & Class        & Notes \\
%\\
00057+4021 & 0.0447 & 11.30 & 1.57 & 1.90 & \nodata\phn & 1 & 7.4(\phn0.5) & S2(1)  & 2 \\
00335$-$2732 & 0.0693 & 11.65 & 2.06 & 2.61 & \nodata\phn & 2 &11.3(\phn0.6) & C(1) & 2 \\
00509+1225 & 0.0611 & 11.31 & 1.59 & 1.78 & \nodata\phn & 2 & 8.8(\phn0.6) & S1(2)  & 2 \\
01364$-$1042 & 0.0484 & 11.53 & 1.89 & 2.00 & \nodata\phn & 3 &15.8(\phn0.7) & L(3,6)	& 2 \\
01417+1651 & 0.0274 & 11.32 & 1.60 & 2.65 & 240\phd\phn & 4 & 40.6(\phn1.3) & L(1),H/L(3),H/A(4) & \\
02483+4302 & 0.0514 & 11.47 & 1.81 & 2.49 & \nodata\phn & 5 &1225(36\phd\phn)  & L(1)&2 \\
03056+2034 & 0.0272 & 10.90 & 1.02 & 1.28 & \nodata\phn & 2 & 17.9(\phn1.0) & H(1)	& 2 \\
03260$-$1422 & 0.0425 & 11.16 & 1.38 & 1.99 & \nodata\phn & 5 &11.4(\phn0.6) & H(1)	& 2 \\
\enddata
\end{deluxetable}

\begin{deluxetable}{rccccclc}
%\scriptsize
\tablecaption{Principal Component Analysis of OHM and Host Properties.
\label{PC_1}}
\tablewidth{0pt}
\tablehead{
\\
& \multicolumn{5}{c}{\bf Correlation Matrix} \\ \relax\\[-1.7ex]
& $\log(f_{100}/f_{60})$ & $\log L_{60\mu m}$ & $\log \Delta v$ 
& $\log L_{OH}$ & $\log L_{1.4GHz}$ \\ \relax\\[-1.7ex] \cline{2-6} \relax\\[-1.7ex]
$\log(f_{100}/f_{60})$ 
& 1.00 & $-$0.37\phs&$-$0.10\phs&$-$0.26\phs&$-$0.14\phs\\
$\log L_{60\mu m}$
&         & 1.00 & 0.38 & 0.49 & 0.73 \\
$\log \Delta v$
&  &  & 1.00 & 0.65 & 0.52 \\
$\log L_{OH}$
&  & & & 1.00 & 0.63 \\
$\log L_{1.4GHz}$
&  & & & & 1.00 \\ \cline{2-6}
\\
Mean &      0.12 & 11.69\phn & 2.15 & 2.87 & 23.52\phn & \boldmath{$N=34$}\\
Std. Dev. & 0.13 &  0.29 & 0.36 & 0.53 &  0.32 \\
\\
& \multicolumn{5}{c}{\bf Eigenvectors} & \multicolumn{2}{c}{\bf Eigenvalues}\\ 
\boldmath{$\hat{e}_1$}& 0.23 &$-$0.48\phs&$-$0.44\phs&$-$0.50\phs&$-$0.51\phs
&$\lambda_1 = 2.81$ & $(\phn56\%)$\\
\boldmath{$\hat{e}_2$}& 0.87 &$-$0.24\phs& 0.39 & 0.15 & 0.14 
&$\lambda_2 = 0.99$ & $(\phn76\%)$\\
\boldmath{$\hat{e}_3$}& 0.36 & 0.53 &$-$0.50\phs&$-$0.36\phs& 0.45 
&$\lambda_3 = 0.67$ & $(\phn89\%)$\\
\boldmath{$\hat{e}_4$}&$-$0.05\phs& 0.28 & 0.63 &$-$0.70\phs& $-$0.15\phs
&$\lambda_4 = 0.33$ & $(\phn96\%)$\\
\boldmath{$\hat{e}_5$}&$-$0.25\phs&$-$0.59\phs& 0.06 &$-$0.32\phs& 0.70 
&$\lambda_5 = 0.20$ & $(100\%)$\\
\\
& \multicolumn{5}{c}{\bf RMS Residuals} \\ 
Principal Axis:    & 0.12 & 0.17 & 0.24 & 0.29 & 0.16 \\
Principal Plane:   & 0.04 & 0.15 & 0.20 & 0.27 & 0.16 \\
Princ. 3-Plane: & 0.02 & 0.09 & 0.13 & 0.23 & 0.10 \\
Princ. 4-Plane: & 0.01 & 0.08 & 0.01 & 0.07 & 0.10 \\
Princ. 5-Plane: & 0.00 & 0.00 & 0.00 & 0.00 & 0.00 \\
\\
& \multicolumn{5}{c}{\bf Observation Errors} \\ 
& 0.08 & 0.04 & 0.06 & 0.05 & 0.04 \\
}
\startdata
\enddata
\end{deluxetable}
\clearpage

%\begin{figure}[!t]
%%\centerline{\psfig{figure=Darling.fig4.epsi,clip=t,width=6in}}
%\caption[OH-FIR-Radio Relations of the Known OH Megamasers]{
%OH-FIR-Radio relations of the known OH megamasers.  
%Available ranges for $L_{FIR}$ are indicated by error
%bars.  
%Dotted lines are fits to the data labeled by their slopes except for
%the $L_{1.4GHz}$ versus $L_{60\mu m}$ plot which is {\it not} at fit
%to the data but instead 
%shows the radio-FIR relation obtained from a large sample of
%galaxies spanning many decades in luminosity (Yun, Reddy \& Condon 2001).
%Shown are 
%{\bf (a)} the OH-FIR relation, 
%{\bf (b)} the OH-60\micron\ relation, 
%{\bf (c)} the radio-60\micron\ relation, and
%{\bf (d)} the unbiased OH-60\micron-radio relation.
%\label{relations:known}}
%\end{figure}

\end{document}